\documentclass[twocolumn, floatfix]{aastex63}
\usepackage[utf8]{inputenc}
\usepackage[T1]{fontenc}
\usepackage[english]{babel}
\usepackage{graphicx}
\usepackage{color}

\usepackage{amsmath, amssymb}    
\usepackage{multirow, booktabs}  

\usepackage{makecell}   

\newcommand{\app}{\texttt{Athena++}}
\newcommand{\athena}{\texttt{Athena}}
\newcommand{\ak}{\texttt{AthenaK}}
\newcommand{\kokkos}{\texttt{Kokkos}}
\newcommand{\parth}{\texttt{Parthenon}}

\definecolor{orange}{rgb}{1.0, 0.5, 0.0}

\shorttitle{AthenaK}
\shortauthors{Stone et al.}
\begin{document}

\title{\ak{}: A Performance-Portable Version of the \app{} AMR Framework}

\author[0000-0001-5603-1832]{James M. Stone}
\affiliation{School of Natural Sciences, Institute for Advanced Study, 1 Einstein Drive, Princeton, NJ 08540, USA}
\affiliation{Department of Astrophysical Sciences, Princeton University, Princeton, NJ 08544, USA}

\author[0000-0003-2131-4634]{Patrick D. Mullen}
\affiliation{CCS-2, Los Alamos National Laboratory, Los Alamos, NM}
\affiliation{Center for Theoretical Astrophysics, Los Alamos National Laboratory, Los Alamos, NM}

\author[0000-0003-3806-8548]{Drummond Fielding}
\affiliation{Department of Astronomy, Cornell University, Ithaca, NY 14853, USA}
\affiliation{Center for Computational Astrophysics, Flatiron Institute, 162 5th Avenue, New York, NY 10010, USA}

\author[0000-0003-3555-9886]{Philipp Grete}
\affiliation{Hamburger Sternwarte, Universit\"at Hamburg, Gojenbergsweg 112, 21029, Hamburg, Germany}

\author[0000-0002-3680-5420]{Minghao Guo}
\affiliation{Department of Astrophysical Sciences, Princeton University, Princeton, NJ 08544, USA}

\author{Philipp Kempski}
\affiliation{Department of Astrophysical Sciences, Princeton University, Princeton, NJ 08544, USA}

\author[0000-0002-0491-1210]{Elias R. Most}
\affiliation{TAPIR, Mailcode 350-17, California Institute of Technology, Pasadena, CA 91125, USA}
\affiliation{Walter Burke Institute for Theoretical Physics, California Institute of Technology, Pasadena, CA 91125, USA}

\author{Christopher J. White}
\affiliation{Center for Computational Astrophysics, Flatiron Institute, 162 5th Avenue, New York, NY 10010, USA}
\affiliation{Department of Astrophysical Sciences, Princeton University, Princeton, NJ 08544, USA}

\author[0000-0001-6952-2147]{George~N.~Wong}
\affiliation{School of Natural Sciences, Institute for Advanced Study, 1 Einstein Drive, Princeton, NJ 08540, USA}
\affiliation{Princeton Gravity Initiative, Princeton University, Princeton NJ 08544, USA}

\correspondingauthor{James M. Stone}
\email{jmstone@ias.edu}

\begin{abstract}
We describe \ak{}: a new implementation of the \app{} block-based adaptive mesh refinement (AMR) framework using the \kokkos{} programming model. Finite volume methods for Newtonian, special relativistic (SR), and general relativistic (GR) hydrodynamics and magnetohydrodynamics (MHD), and GR-radiation hydrodynamics and MHD, as well as a module for evolving Lagrangian tracer or charged test particles (e.g., cosmic rays) are implemented using the framework. In two companion papers we describe (1) a new solver for the Einstein equations based on the Z4c formalism and (2) a GRMHD solver in dynamical spacetimes also implemented using the framework, enabling new applications in numerical relativity.  By adopting \kokkos{}, the code can be run on virtually any hardware, including CPUs, GPUs from multiple vendors, and emerging ARM processors. \ak{} shows excellent performance and weak scaling, achieving over one billion cell updates per second for hydrodynamics in three-dimensions on a single NVIDIA Grace Hopper processor and with a typical parallel efficiency of 80\% on 65536 AMD GPUs on the OLCF Frontier system. Such performance portability enables \ak{} to leverage modern exascale computing systems for challenging applications in astrophysical fluid dynamics, numerical relativity, and multimessenger astrophysics.
\end{abstract}


\section{Introduction}

Given the critical role that numerical methods play in theoretical astrophysics, it is important that software that implements such methods be able to exploit
the most capable high-performance computing (HPC) systems available.
In recent years, however, HPC systems have become increasingly heterogeneous, incorporating novel architectures such as ARM processors
or external accelerators such as graphical processing units (GPUs).  Programming such systems often requires adopting vendor-specific language extensions such as CUDA (from NVIDIA), HIP (from AMD), or SYCL (from Intel) or relying on the vendor to support open source standards such as OpenACC or OpenMP.  Developing and maintaining software that runs effectively across diverse platforms is a challenge.

This issue is not confined to astrophysics, and fortunately generic solutions are emerging from the scientific computing community at large.  For example, the \kokkos{}\footnote{https://kokkos.org/} programming model \citep{9485033} is one such solution. It consists of C++ templates and abstractions that can be used
to manage execution on heterogeneous systems with multiple memory spaces, execution spaces, and execution policies (e.g., scalar, vector, or shared-memory parallelism). When building software written with \kokkos{} for a particular HPC system, instructions are effectively translated into the native programming environment of the target architecture, enabling vendor optimizations that can improve performance.  Thus, software developed with \kokkos{} becomes {\em performance-portable}.

This paper describes \ak{}: a completely new implementation of the adaptive mesh refinement (AMR) framework, fluid dynamics solvers, and general relativistic radiation transport solver in the 
\app{} code \citep[hereafter S20]{Stone2020ApJS..249....4S} using the \kokkos{} programming model.
In \ak{} the AMR and mesh are abstracted from the physics so that they can be used to solve a diverse range of problems.  In two companion papers we describe a new solver for the Einstein equations of general relativity based on the Z4c formalism that has been implemented using the \ak{} AMR framework \citep{Zhu.9.24}, as well as a new GRMHD solver that 
works in dynamical spacetimes \citep{Fields.9.24}.
We show that the AMR framework and solvers in \ak{} achieve excellent performance and parallel scaling across a wide range of 
architectures, including both CPUs and GPUs from a variety of vendors.
Taken together these solvers enable problems in astrophysical fluid dynamics, numerical relativity, and multimessenger astrophysics to be tackled using modern exascale HPC systems.

\ak{} is not the first nor the only astrophysical fluids code that adopts \kokkos{}.  In fact, \cite{grete2021} have already described an implementation of the MHD solvers in \app{} on a uniform mesh using Kokkos. Subsequently the block-based AMR framework in \app{} has also been independently implemented in the \parth{} framework \citep{grete2022}, and the latter now serves as the foundation for an increasing number of astrophysical application codes including \texttt{AthenaPK}\footnote{https://github.com/parthenon-hpc-lab/athenapk}, \texttt{KHARMA} \citep[described in][]{prather2024}, and \texttt{Phoebus} \citep{2023AAS...24110547R}.  Recently, \cite{lesur2023} have developed \texttt{IDEFIX}, a new implementation of the \texttt{PLUTO} code \citep{mignone2012ApJS..198....7M} in \kokkos{}.

The \ak{} code is in fact an outgrowth of the \parth{} collaboration, and it shares many features in common both with the \parth{} framework and downstream codes such as \texttt{AthenaPK}, although there are also important differences. \parth{} is a much more ambitious, complex, and capable framework that has many extensions that will make it useful for a wide variety of applications across many different disciplines. The \ak{} code described in this paper is a much smaller and simpler project, focused on applications in astrophysical fluid dynamics. 

There are a variety of other astrophysical fluid codes that run on GPU-accelerated hardware by adopting vendor-specific tools such as CUDA.  These include an implementation of the \texttt{RAMSES} code in CUDA \citep{kestener2017ascl.soft10013K}, \texttt{Cholla} \citep{schneider2015ApJS..217...24S} and more recently \texttt{Cholla-MHD} \citep{caddy2024ApJ...970...44C}, \texttt{H-AMR} \citep{liska2022ApJS..263...26L}, and \texttt{GAMER-2} \citep{schive2018MNRAS.481.4815S}.  Other block-structured AMR packages that are GPU-enabled include \texttt{AMReX} \citep{zhang2019JOSS....4.1370Z}.

The \ak{} code described in this paper should be thought of as complementary to these other codes.
What distinguishes it is the combination of (1) performance portability through \kokkos{}, (2) full block-structured AMR framework, (3) a wide variety of physics solvers, and (4) open source development which builds on previous versions of Athena \citep[hereafter S08]{Stone+2008}.  Users of previous versions of \texttt{Athena} should be able to adopt their applications to \ak{} very easily.

In the following section, we describe the modifications necessary to
implement the \app{} AMR framework in \ak{} using \kokkos{}.  In \S3 we describe the various fluid dynamic and GR-radiation transport solvers implemented in \ak{}, while in \S4 we describe a particle module that includes methods for both plasma kinetics and Lagrangian tracer particles. In \S5 we provide performance and scaling data on various architectures for the fluid solvers that demonstrate true performance portability.  Finally, in \S6 we conclude.

\section{AMR Framework}

There are several inherent complexities associated with programming heterogeneous computing systems.  Firstly, data may be stored in multiple memory spaces (e.g., on the host, on the device, or in a shared location), and the programmer is responsible for deciding what goes where.  Similarly there may be multiple execution spaces (e.g., CPUs on a host as well as external accelerators such as GPUs), and again the programmer must control what is computed where.  Finally, each different execution space may have a different architectures, i.e. scalar, vector, or shared-memory parallel processors, and the programmer must be able to control the execution strategy in each space individually.  

Given these complexities, we found it easier to rewrite the \app{} AMR framework from scratch rather than trying to port it to \ak{}.  In the end, very little of the overall design was changed.  Thus, \ak{} adopts the same block-based AMR design as in \app{} with communication between levels occurring only at boundaries, flux corrections at level boundaries to enforce conservation, uniform time stepping across levels, and dynamical execution via a task list.  Section 2 of S20 continues to serve as the primary reference for this framework; in the paragraphs below we only describe the extensions and modifications required for good performance across different architectures.  We note that many of these ideas originate from the \parth{} collaboration, and therefore replicate features of that framework \citep{grete2022}.

{\bf \texttt{MeshBlockPack}s:} As in \app{}, in \ak{} the mesh is divided into structures termed \texttt{MeshBlock}s which are organized into a binary tree (octree in three dimensions) and labeled via Z-ordering.  However, in \ak{} we have found it important to organize all \texttt{MeshBlock}s on the same device into an aggregate data structure called a \texttt{MeshBlockPack}.  Since the operations in all \texttt{MeshBlock}s on the same device can be performed in parallel, this reduces the number of kernel launches and greatly enhances performance when there are a large number of small \texttt{MeshBlock}s on the same device.

{\bf Physics modules:} Each physics module is stored as a \texttt{C++} object in each {\tt MeshBlockPack}. Each module is independent with its own memory and task list. This reduces the combinatorial complexity of building task lists, and enables multiple physics objects to be constructed simultaneously for calculations involving multiple fluids or numerical relativity with fluid dynamics. 

{\bf Data locality:} Dependent variables within each physics module are stored as $(N+1)$-dimensional \kokkos{} {\tt View}s, with the extra dimension being the index of the \texttt{MeshBlock} associated with the data. This data is always stored on the device, and is never moved to the host except for outputs. Conversely, the mesh binary tree is created, stored, and refined (with AMR) on the host. Any mesh data that needs to be accessed on both the host and device are stored as \kokkos{} {\tt DualView}s. Fortunately the syntax of {\tt View}s is nearly identical to the multidimensional array class in \app{}, making the transition seamless.

{\bf Execution spaces:} Operations on dependent variables are always performed on the device, with the host controlling the creation and manipulation of the tree, and managing dynamic execution via the task list. To the extent possible, the default memory and execution spaces for any computationally intensive operations in \ak{} are always on the device.

{\bf Boundary conditions:} As in both \athena{} and \app{}, boundary conditions for the dependent variables on each \texttt{MeshBlock} are applied via ghost zones. Communication of data between \texttt{MeshBlock}s to load the ghost cells is done within parallel kernels on the device, either as simple memory copies or to load data into a boundary buffer (allocated in device memory) that is then communicated via point-to-point \texttt{MPI} calls. The number of \texttt{MPI} messages sent between ranks is proportional to the number of \texttt{MeshBlock}s stored within the \texttt{MeshBlockPack} on each rank.  When communicating data between different levels in an AMR grid, prolongation and restriction operations are also performed all at once inside parallel kernels.  The boundary communication infrastructure in \ak{} has been largely redesigned from the original version in \app{} to make it simpler and more flexible on shared-memory parallel devices.

{\bf Cartesian coordinates only:} Currently, \ak{} only supports uniform Cartesian coordinates for a number of reasons.  Firstly, applications with \ak{} target exascale HPC systems that enable very large calculations, and such calculations are generally best discretized with Cartesian coordinates.  Moreover the difference between calculations performed in different coordinates is only at the level of truncation error, which should be small for any resolved calculation.  In fact higher-order spatial discretizations are simpler in Cartesian coordinates, which can make them more accurate than curvilinear coordinates. Finally many elements of object-orientated programming such as virtual functions and polymorphism are not supported by the compilers currently available on many GPUs, and this means implementing different coordinates using the design in \app{} would require significant changes. The primary limitation of not supporting curvilinear coordinates is that 2D axisymmetric calculations are not possible with \ak{}.  Any applications that truly require curvilinear coordinates can always adopt \app{} or other codes.

{\bf Generalized task list:} The design of the task list in \ak{} has been improved to support more general operations.  Separate task lists are built for multi-stage time integrators, operator split source terms, and multiple physics modules and then chained together, or interleaved as necessary. Which physics is included in a calculation is controlled simply by which set of tasks (and task lists) are executed at run time.

{\bf I/O:} Data required for output is generally stored on the device, and so must first be copied to the host for output. However, this enables asynchronous outputs since the device can modify its own data while the host continues to write output. Parallel files are written using \texttt{MPI-IO} to a custom binary format with no dependence on external libraries.  \texttt{Python} scripts are provided that can translate \ak{} binary files to other formats, such as \texttt{HDF5}.

{\bf Building and compiling:} \ak{} adopts {\tt cmake} to control building the code for different architectures, largely because \kokkos{} supports external builds using this tool.  This allows the numerous features of \kokkos{} to be controlled from command-line options during the build step. The physics and algorithm options in \ak{} are no longer controlled by a configure script as in \app{}.  Instead, the entire source code is compiled and options are selected at run time using parameters in the input file, including a large number of problem generators (i.e., the functions that generate the initial conditions for a wide range of problems). This greatly simplifies testing and development. User-defined problem generators are easily included as a \texttt{cmake} command line option. 
 Since the number of registers is extremely limited on many GPUs, we have found that the use of templating for some algorithmic options (e.g., Riemann solvers) reduces register spillage and can greatly improve performance.

{\bf Supporting open source development:} The entire \ak{} source code is available in an public repository\footnote{\tt{https://github.com/IAS-Astrophysics/athenak}}, there are no private development versions. The code continues to be licensed under the permissive BSD-3 open source license.  Contributions and feedback from the community are encouraged.

\section{Fluid Solvers}

Numerical methods for Newtonian, special relativistic (SR), and general relativistic (GR) hydrodynamics and MHD, and GR-radiation hydrodynamics and MHD, have all been implemented in \ak{}. In this section we describe each of these solvers, while in \S5 we provide performance metrics on various devices.

The fluid solvers all integrate the coupled system of conservation laws 
\begin{equation}
\frac{\partial {\bf U}}{\partial t} + \nabla \cdot {\bf F}({\bf W}) = \bf{S}
\label{eq:fluid}
\end{equation}
where $\bf{U}$ and $\bf{F}$ are vectors of the conserved variables and their fluxes respectively, while $\bf{W}$ is a vector of primitive variables and $\bf{S}$ are source terms (if required).  In general, an equation of state which relates the various
thermodynamic variables must be specified to close the system.

For MHD, the induction equation for the evolution of the magnetic field ${\bf B}$ must be evolved
\begin{equation}
\frac{\partial {\bf B}}{\partial t} + \nabla \times {\bf E} = 0
\label{eq:induction}
\end{equation}
where ${\bf E}$ is the electric field which in general includes both inductive and non-ideal terms.

The numerical algorithms adopted by the fluid solvers are all based on higher-order finite-volume schemes.  These schemes adopt a reconstruct, solve, and advance strategy. The details of the methods and their implementation in the \texttt{Athena} codes have already been published in a series of papers \citep{gs05, gs08, Stone+2008, Beckwith2011, WhiteStone2016, Stone2020ApJS..249....4S, 2023ApJ...949..103W} and will not be repeated here, instead only general features of the implementation that are unique to \ak{} will be described.

For spatial reconstruction schemes, the piecewise linear method (PLM), two versions of the piecewise parabolic method using either the original \citep[hereafter PPM4]{ColellaWoodward1984} or extremum-preserving limiters \citep[hereafter PPMX]{McCorquodale2015a}, and the improved WENOZ scheme of \cite{ACKER2016726} are implemented. These are selected by a parameter in the input file.  Reconstruction is always performed on the primitive variables.  As shown in S08, it is often advantageous to reconstruct the characteristic variables, however this option has not yet been implemented in \ak{}.

To compute fluxes of the conserved variables at cell interfaces, a variety of different Riemann solvers \citep{toro2013riemann} are implemented depending on the physics, including the local Lax-Friedrichs (LLF) method, various versions of the Harten-Lax-van Leer method including HLLE, HLLC, and HLLD, as well as Roe's linearized solver.

To time-advance the solution, the method-of-lines approach is adopted.
A variety of low storage, high-order strong-stability-preserving (SSP) Runge-Kutta (RK) integration
schemes \citep{Ketcheson2010} are implemented in \ak{}, from RK1 through RK4. Explicit time intergration requires the time step satisfy the constraint
\begin{equation}
    \Delta t \leq \mathrm{CFL} \cdot \min \left[ (\Delta x_i)/{\lambda}_{{\rm max},i} \right]
\end{equation}
where ${\lambda}_{{\rm max},i}$ is the maximum cell-centered signal speed in the $i$-th dimension in each cell, and the minimum is taken over all cells and all dimensions. The factor $\mathrm{CFL} \leq 1$ depends on the specific time integrator and the dimensions $N_\mathrm{dim}$ of the problem.  Generally we use $\mathrm{CFL} \leq 1 / N_\mathrm{dim}$.
Note that the \athena{} code also used the dimensionally-unsplit CTU algorithm of \cite{Colella1990}, which has many advantages for non-relativistic problems. However, since it does not extend easily to the relativistic equations of motion, a CTU integrator has not been implemented in \ak{}.

For some problems which include very stiff source terms, implicit-explicit (IMEX) RK integrators have proved to be useful \citep{ASCHER1997151}.
However, as shown by \cite{krapp2024ApJS..271....7K}, many commonly used IMEX integrators fail to guarantee positivity of the solution, which is very problematic when stiff source terms are included in the energy equation. In \ak{} we have implemented several different IMEX integrators, and we describe their application to two-fluid MHD in \S3.3.

For MHD, the magnetic field is discretized as area-averages on cell-faces, and the constrained transport (CT) algorithm of \cite{eh1988ApJ...332..659E} is used to preserve the divergence-free constraint with upwinded electric fields calculated using the method of \cite{gs05, gs08}. As discussed in S20, using face-centered fields complicates communication of ghost zones and flux conservation in the AMR framework, however we have found the use of the staggered mesh version of CT is critical for some problems.

Guaranteeing positivity of the density and internal energy in multidimensional higher-order finite-volume schemes is challenging in regions where the kinetic or magnetic energy is very large.  This is not due to round-off error associated with differencing large numbers, but rather truncation error inherent to the scheme can become comparable to the magnitude of derived quantities such as pressure.  To overcome this issue,  in \ak{} we have implemented the first-order flux-correction (FOFC) scheme used in \cite{lemaster2009ApJ...691.1092L}. In each stage of the SSP-RK integrators, if the density or pressure in any cell were to go negative, the fluxes on all faces of that cell are recalculated using first-order (piecewise constant) reconstruction and a diffusive LLF Riemann solver.  This must be performed before flux corrections associated with level boundaries are made when using AMR. In practice, FOFC is applied extremely rarely \citep[once in every $\sim10^{10}$ cell updates in the supersonic turbulence application presented in][]{lemaster2009ApJ...691.1092L}.  Nevertheless we find using FOFC greatly improves the stability of calculations in extreme regimes, for example it significantly reduces the need for floors in general relativistic MHD simulations of black hole accretion flows (see \S3.5 and \S3.6).

\subsection{Non-relativistic Hydrodynamics}

\begin{figure*}[htb]
    \centering
    \includegraphics[width=3.5in, keepaspectratio=true]{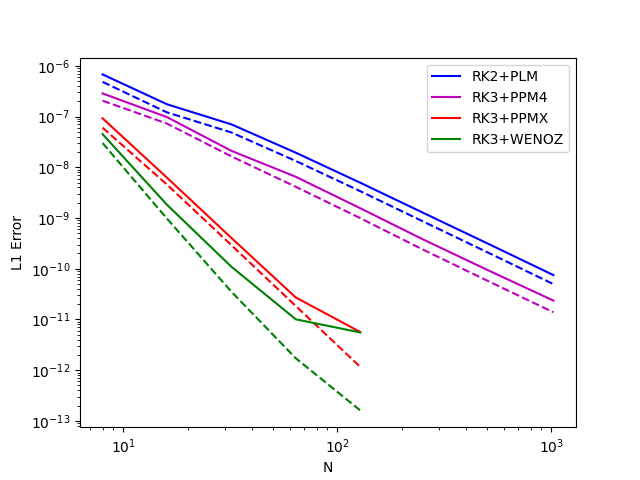}
    \includegraphics[width=3.5in, keepaspectratio=true]{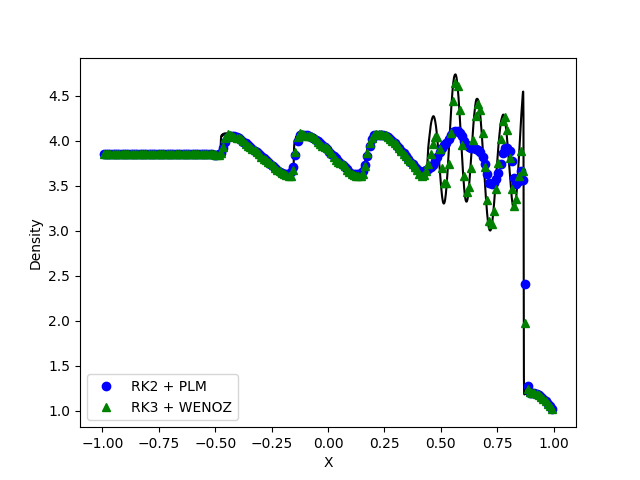}
    \caption{(Left) Linear wave convergence rates in 1D for both entropy (solid-line) and sound (dashed-line) waves for different methods.  (Right) Density in the Shu-Osher shocktube test using 200 uniform cells and different methods.  The solid black line is a reference solution computed using 2000 cells and RK3+WENOZ.}
    \label{fig:Hydro_tests}
\end{figure*}

\begin{figure*}[htb]
    \centering
    \includegraphics[width=\linewidth]{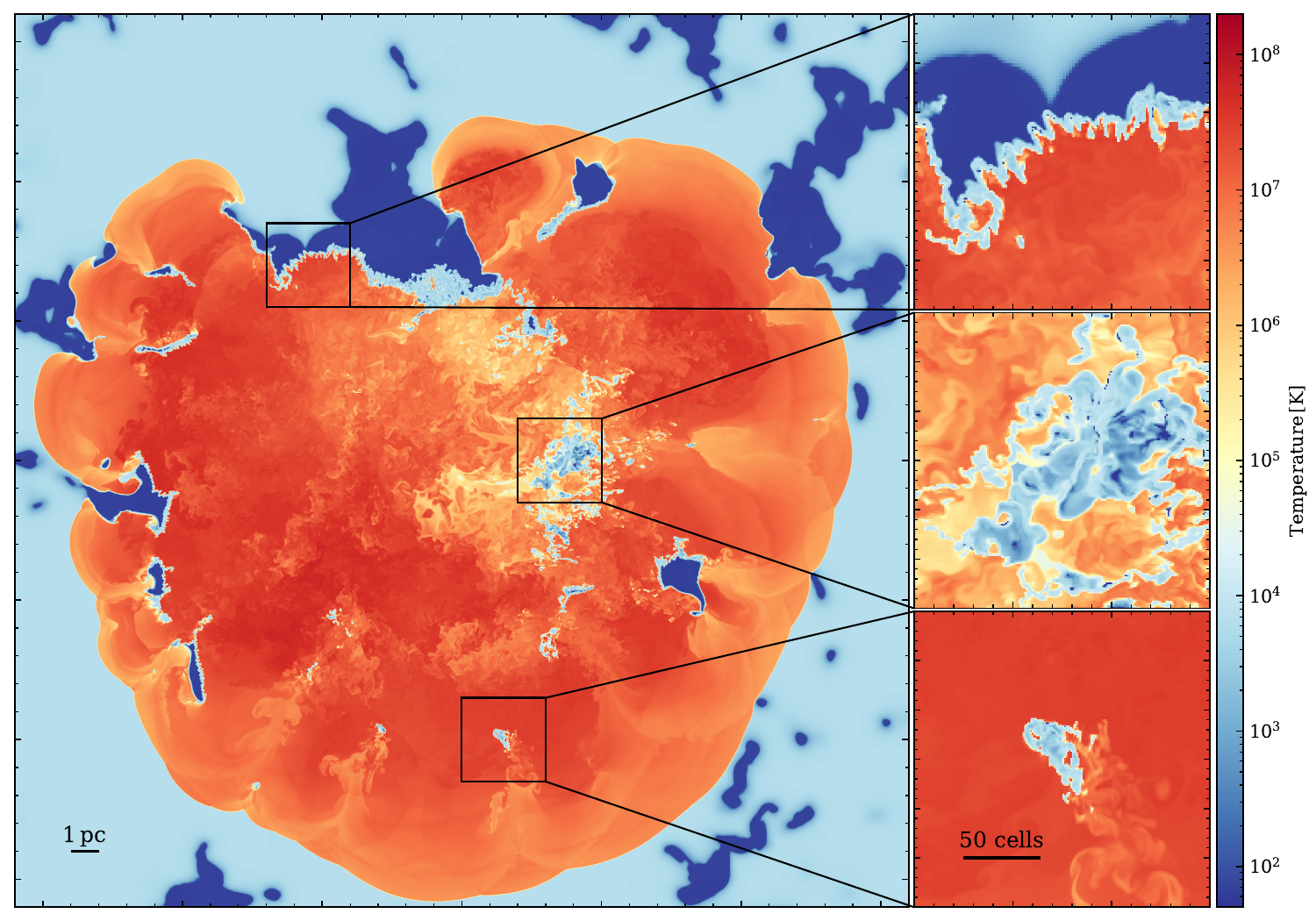}
    \caption{Slice of temperature through the $z=0$ plane at time $t=10\,\mathrm{kyr}$ in a 3D simulation of a supernova blast wave interacting with a turbulent multiphase medium on a $2048^3$ grid. Zoom-ins of selected regions in left panels show nonlinear thin-shell instability (top), turbulent mixing (middle), and single shock-cloud interaction (bottom). }
    \label{fig:Hydro_app}
\end{figure*}

In Newtonian hydrodynamics the conserved variables and their fluxes are
\begin{align}
{\bf U} &= \begin{bmatrix}
           \rho \\
           \rho {\bf v} \\
           E 
         \end{bmatrix},
\hspace*{0.1cm}{\bf F} = \begin{bmatrix}
           \rho {\bf v} \\
           \rho {\bf v}{\bf v} + p_g \\
           (E+p_g) {\bf v}
         \end{bmatrix},
\label{eq:hydro_vars}
\end{align}
where $\rho$ is the mass density, ${\bf v}$ the fluid velocity, $p_g$ the gas pressure, $E=u_g + \rho v^2/2$ the gas total energy, and $u_g$ the gas internal energy. The vector of primitive variables is ${\bf W} = \left[ \rho, {\bf v}, u_g \right]$.  An equation of state (EOS) $p_g=p_g(\rho,u_g)$ is required to close the system. For non-relativistic hydrodynamics in \ak{}, both a $\gamma$-law EOS $p_g = (\gamma - 1) u_g$ and an isothermal EOS $p_g = C^2 \rho$ are implemented, where $\gamma$ is the ratio of specific heats and $C$ the isothermal sound speed respectively.  In the latter case the energy equation in (\ref{eq:fluid}) is not needed.  

The accuracy and fidelity of the various reconstruction schemes, Riemann solvers, and time-integration algorithms implemented in \ak{} have been documented in many other references \citep[e.g. S20,][]{toro2013riemann}.  Here, to demonstrate their implementation in \ak{}, we only present results from two simple test problems.  Both are run on a single CPU core.

The first, and most quantitative, is to measure the convergence of errors in the propagation of linear waves.  The left panel of Figure \ref{fig:Hydro_tests} shows the convergence of the $L_1$ error for both sound and entropy waves in one-dimensional hydrodynamics using various algorithmic options.  Following S20, the problem is initialized using an exact eigenmode for each wave family and evolved for one wave period with a variety of spatial resolutions.  In one-dimension, fourth-order convergence is observed for the extremum-preserving reconstruction schemes PPMX and WENOZ until either non-linear effects or round-off error becomes important, while both PPM with more aggressive limiters and PLM show second-order convergence.

The second test is the Shu-Osher shocktube \citep{shu1989JCoPh..83...32S} which involves the interaction of a discontinuity with a smoothly varying background, and therefore demonstrates both the shock-capturing and diffusive properties of different methods. The right panel of Figure \ref{fig:Hydro_tests} shows the density at $t=0.047$ when run with both PLM+RK2 and WENOZ+RK3 with 200 uniform grid cells.  Clearly the extremum-preserving higher-order WENOZ method captures small scales in the problem more accurately than PLM.

Finally, to showcase how \ak{} can enable new applications at scale, we show preliminary results from a challenging calculation involving the interaction of a supernova blast wave with a turbulent, multiphase medium. Both heating and cooling source terms representative of physical processes in the ISM \citep{Schneider2018ApJ...860..135S} are included in the energy equation. The calculation begins with an initially uniform thermally unstable background in a cubic domain of size $32^3$~pc on a grid of $2048^3$ cells to which small perturbations are added. The system is evolved until the thermal instability has gone nonlinear, generating a turbulent multiphase medium, at which point a spherical blast wave is initialized in the center of the domain by adding a region of high pressure. As this blast wave propagates through the background, it encounters dense cold filaments and clumps which produce further turbulence and mixing. The simulation uses an RK3+PPM4 integrator, the HLLC Riemann solver, and FOFC. Figure \ref{fig:Hydro_app} shows a slice of the temperature at late time. Insets in the figure show specific regions of interest where nonlinear thin-shell instability, turbulent mixing, and shock-cloud interaction are occurring. The full calculation took only $2.4\times10^3$ GPU hours. A full analysis of the results, including how the time evolution of the shock wave is affected by mass entrainment and mixing is given in \cite{Guo.9.24}.

\subsection{Non-relativistic MHD}

\begin{figure*}[htb]
    \centering
    \includegraphics[width=3.2in, keepaspectratio=true]{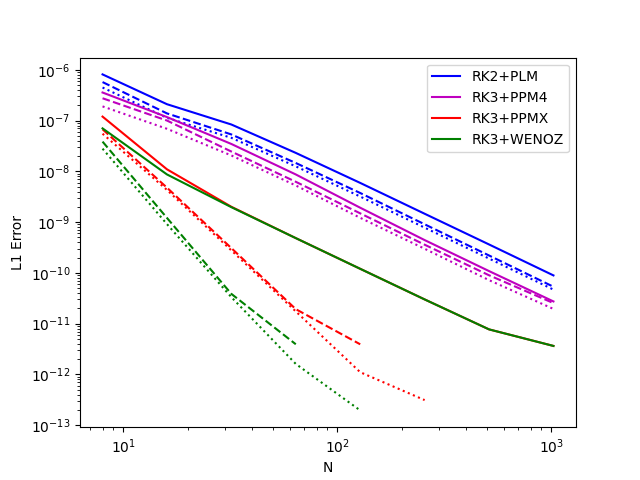}
    \includegraphics[width=3.2in, keepaspectratio=true]{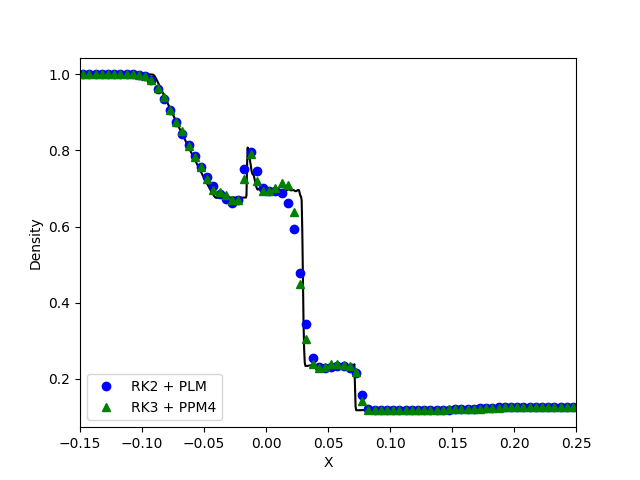}
    \caption{(Left) MHD linear wave convergence rates in 1D for fast (solid-line), slow (dashed-line), and Alfv\'{e}n (dotted-line) waves for different methods.  (Right) Density in the Brio-Wu shocktube test using 200 uniformally-spaced cells and different methods.  The solid black line is a reference solution computed using 2000 cells and RK3+PPM4.}
    \label{fig:MHD_tests}
\end{figure*}

In Newtonian MHD the conserved variables and their fluxes are
\begin{align}
{\bf U} &= \begin{bmatrix}
           \rho \\
           \rho {\bf v} \\
           E 
         \end{bmatrix},
\hspace*{0.1cm}{\bf F} = \begin{bmatrix}
           \rho {\bf v} \\
           \rho {\bf v}{\bf v} - {\bf B} {\bf B} + p^* \\
           (E+p^*) {\bf v} - {\bf B} \left({\bf B} \cdot {\bf v} \right)
         \end{bmatrix},
\label{eq:mhd_vars}
\end{align}
where $p^* = p_g + p_m$ is the sum of gas and magnetic pressures. For non-relativistic MHD, both a $\gamma$-law EOS (with $E=p_g/(\gamma - 1) + \rho v^2/2 + B^{2}/2$) and an isothermal gas EOS (with $p_g=C^2 \rho$) are implemented.  

The left panel of Figure \ref{fig:MHD_tests} shows the results for a one-dimensional linear wave convergence test for each MHD wave family using various algorithmic options run on a single CPU core.  While fourth-order convergence is observed for RK3+PPMX and RK3+WENOZ for slow and Alfv\'{e}n waves, the fast wave only converges at second-order even in one-dimension.  For PPM4 and PLM reconstruction, all waves converge at second-order, as expected.  

The right panel of Figure \ref{fig:MHD_tests} shows the density in the Brio-Wu shocktube \citep{BrioWu} using different methods run on a CPU. Since the solution mostly involves discontinuities, both methods perform similarly.  Note the small amplitude ringing near the compound wave at $x=0$.  This can be significantly reduced by reconstructing characteristic variables instead of primitive variables (S20), although this option (which is complex in SR and GR) has not been implemented in \ak{}.

\begin{figure*}[htb]
    \centering
    \includegraphics[width=\linewidth]{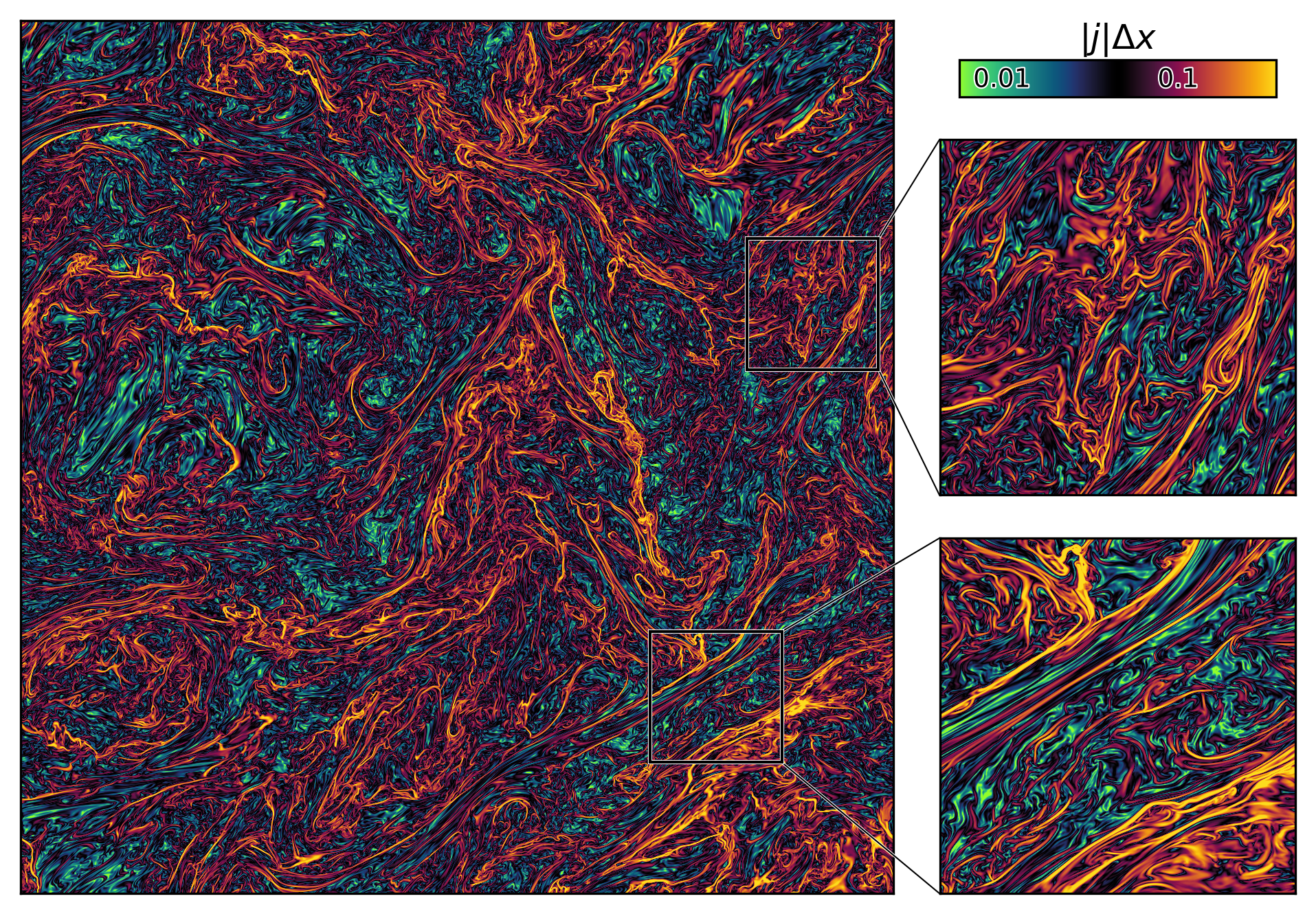}
    \caption{Magnitude of the current density in a 3D simulation of driven MHD turbulence at a resolution of $4096^3$. Insets show zoom-ins to selected regions.}
    \label{fig:MHD_app}
\end{figure*}

Finally, Figure \ref{fig:MHD_app} shows a simulation modeling isothermal, driven MHD turbulence on a $4096^3$ grid including explicit viscosity and resistivity (see \S\ref{sec:diffusion}) at a Reynolds number of $2 \times 10^5$ and a magnetic Prandtl number of 1. The simulation uses WENOZ for spatial reconstruction, RK2 for time integration, an HLLD Riemann solver, and was run on 4096 GPUs on OLCF's Frontier using $256^3$ {\tt MeshBlock}s. Turbulence is driven only on large scales, with an acceleration field following a parabolic power spectrum between $1 \leq 2 \pi k/L_{\rm box} \leq 3$, peaking at $2 \pi k = 2 L_{\rm box}$. The driving is applied in real space rather than Fourier space to avoid the costly global communications required for FFTs in large domains. The turbulence driving field is restricted to be purely solenoidal (i.e., the driving field is divergence-free). The turbulent energy injection rate is set so that the root-mean-squared velocity is half of the sound speed, yielding $\mathcal{M} = v_{\rm rms} / c_{s} = 0.5$. The mean magnetic flux in the domain corresponds to an Alfv\'{e}n Mach number of 2. After several eddy turnover times, the tangled magnetic field component dominates over the mean field, reaching a saturation ratio of approximately $\delta B / \bar{B} \approx 3$. The simulation was run for 4 eddy turnover times. The plot shows the magnitude of the current normalized by the grid spacing in a 2D slice through the 3D domain. The zoomed-in panels highlight the intricate small-scale structures typical of high-amplitude MHD turbulence.

\subsection{Diffusion}
\label{sec:diffusion}

\begin{figure*}[htb]
    \centering
    \includegraphics[width=\linewidth]{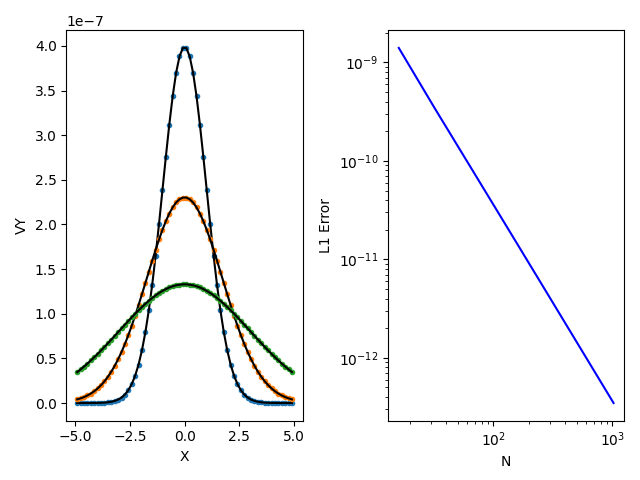}
    \caption{(Left) Viscous diffusion of a Gaussian profile of the transverse velocity $v_y$ at times of 1, 2, and 4 using a 1D mesh of 64 cells.  The analytic solution is given as a solid line.  (Right) $\propto N^{-2}$ convergence of the $L_1$ error norm in this test.}
    \label{fig:Diffusion_test}
\end{figure*}

\begin{figure*}[htb]
    \centering
    \includegraphics[width=3.5in, keepaspectratio=true]{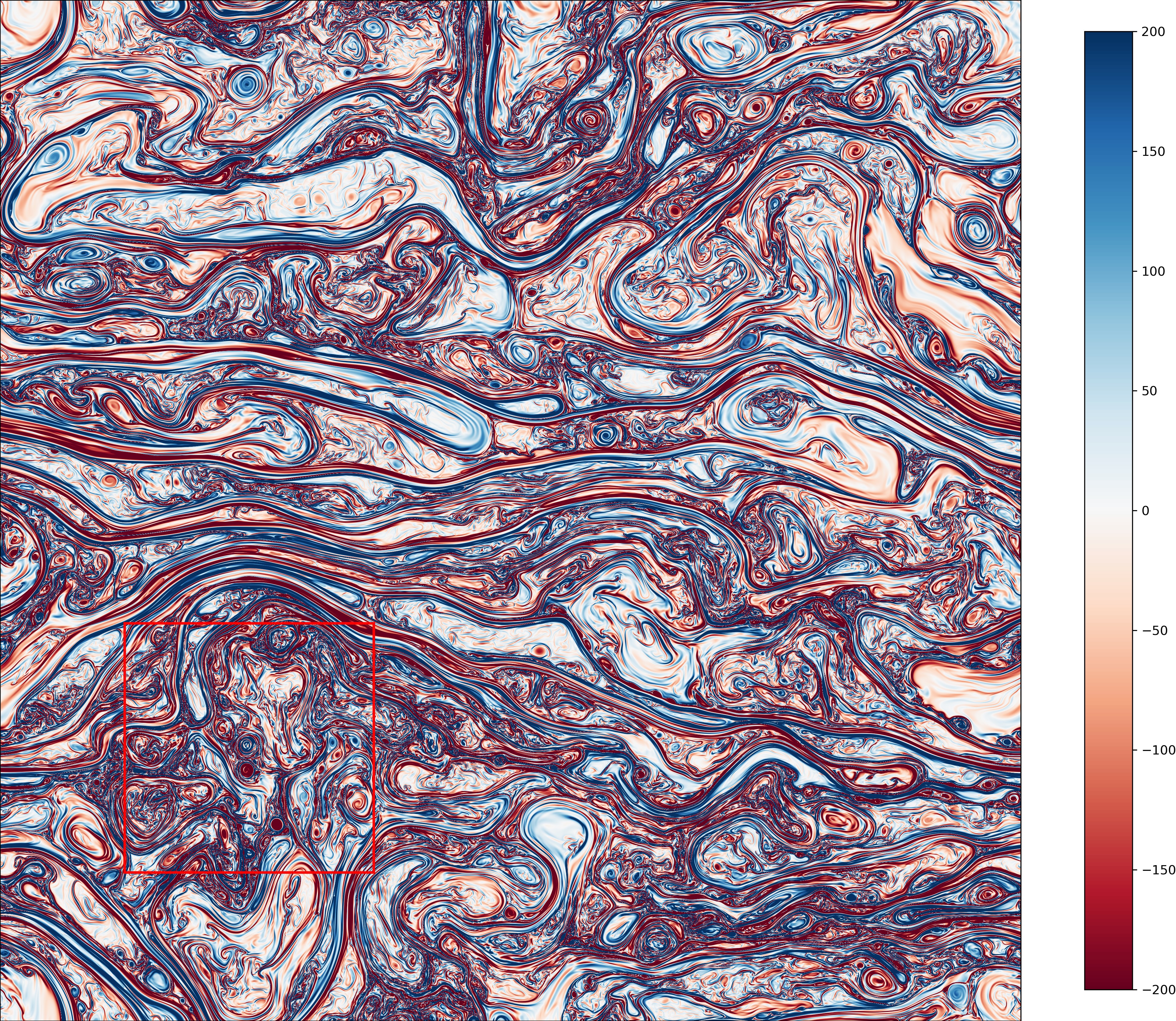}
    \includegraphics[width=3.5in, keepaspectratio=true]{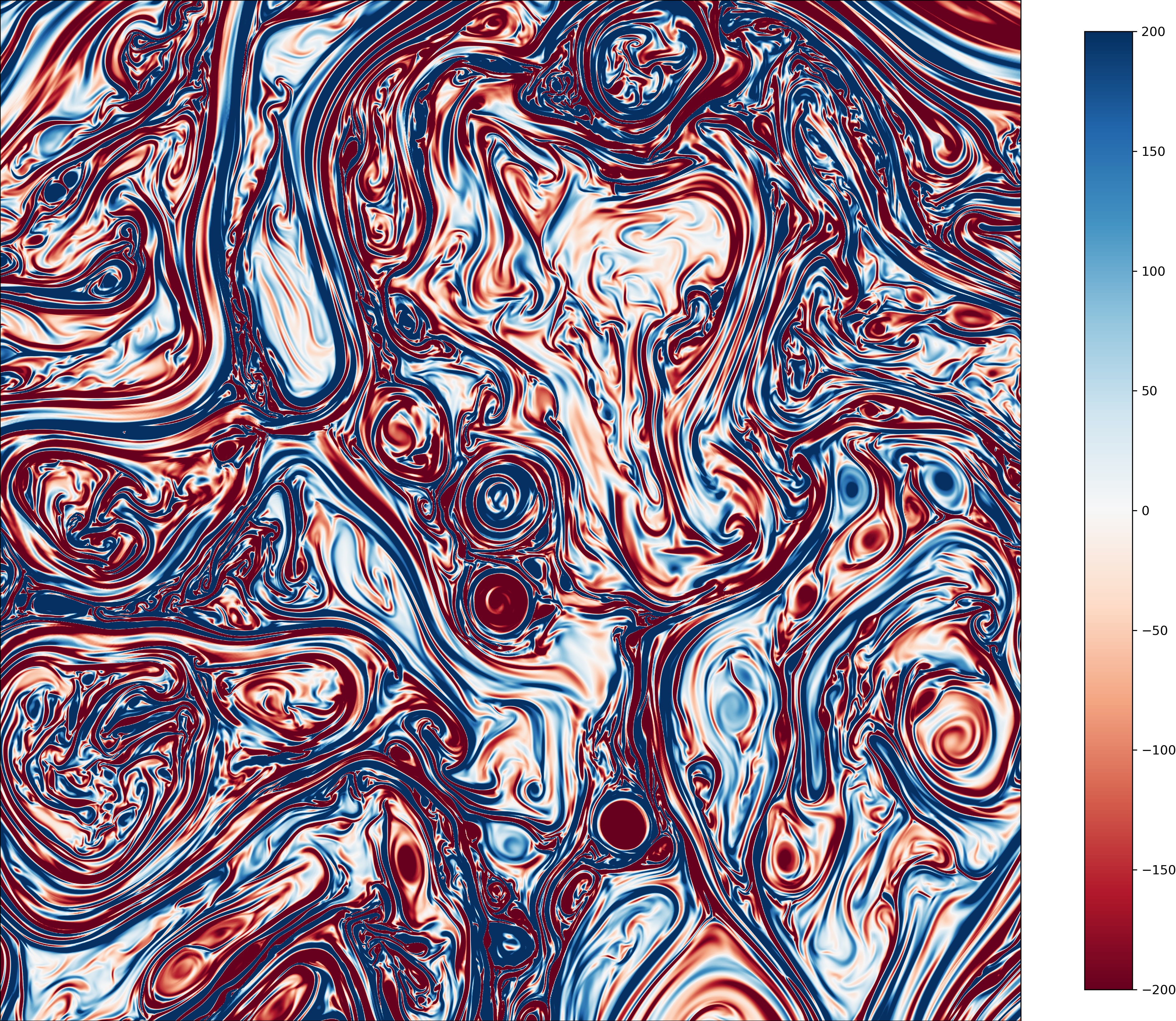}
    \caption{(Left) Out-of-plane component of the current density in a two-dimensional (axisymmetric) simulation of the MRI with no-net vertical flux, and with explicit resistivity and viscosity ($Re = Re_{M} = 10^6$) at extreme resolution. (Right) Zoom-in to the small region outlined in red in left image. Images are shown after evolving for 10 orbits.}
    \label{fig:Diffusion_app}
\end{figure*}

Diffusive processes such as viscosity and thermal conduction can be included in non-relativistic problems through the addition of the appropriate fluxes to ${\bf F}$ in Equation (\ref{eq:fluid})
\begin{align}
{\bf F}_{\rm visc} = \left[
   0, {\bf \Pi},({\bf v} \cdot {\bf \Pi})
   \right]
\label{eq:visc_flux},
\end{align}
\begin{align}
{\bf F}_{\rm cond} = \left[
   0, 0,\kappa \nabla T
   \right]
\label{eq:cond_flux},
\end{align}
where $\mathbf{\Pi}$ is the viscous stress tensor (see Equation 7 in S20).
Ohmic resistivity contributes to the flux
\begin{equation}
    \mathbf{F}_\mathrm{Ohm} = \left[0, 0, \eta \mathbf{J} \times \mathbf{B} \right]
\end{equation}
\textit{and} supplements the electric field ${\bf E}$ in Equation (\ref{eq:induction}) with a non-ideal term
\begin{align}
{\bf E}_{\rm Ohm} = \eta {\bf J}
\label{eq:e_ohm},
\end{align}
where ${\bf J}=(\nabla \times {\bf B})$ is the current density. When added in this fashion, the diffusive fluxes are computed explicitly at every stage of the RK time integrators rather than using operator splitting.

Adding diffusive terms to the evolution equations (\ref{eq:fluid}) makes them of mixed hyperbolic-parabolic form; explicit time integrators are thereby subject to a much more restrictive stability limit,
\begin{equation}
    \Delta t \leq \mathrm{CFL_\mathrm{diff}} \cdot \mathrm{min}(\left[\Delta x_i \right]^2 / \left[2 N_\mathrm{dim} D_i \right])
\end{equation}
where $D$ is the diffusion coefficient of the processes included.  At high resolutions or when $D$ is large this limit may be prohibitive, and generally operator splitting methods combined with an implicit solver or super-time-stepping \citep{Meyer14} are required. While the generalized task list in \ak{} makes implementing these approaches straightforward, as of yet they have not been ported from \app{}.

To demonstrate the inclusion of diffusive terms in \ak{}, the right panel of Figure \ref{fig:Diffusion_test} shows the viscous diffusion of a Gaussian profile of the transverse component of velocity $v_y$ in a 1D uniform grid at three different times, run on a single CPU core.  The coefficient of viscosity $\nu=1$, while the RMS width of the initial Gaussian $\sigma=0.5$, giving a diffusion time of $\sigma^2/\nu = 0.25$.  The analytic solution for this problem is also plotted as a solid line. The $L_1$ errors in the solution plotted in the left panel of the figure show convergence at second-order, as expected for the second-order finite differencing used to compute the viscous stress tensor ${\bf \Pi}$.  If higher-order integration methods are adopted, e.g. \cite{felker2018JCoPh.375.1365F}, then higher-order differencing of the diffusive fluxes must be used as well.

Figure \ref{fig:Diffusion_app} shows an MHD application with explicit viscosity and resistivity, namely the evolution of the magnetorotational instability (MRI) in 2D with no net vertical flux.  With \ak{}, and using only 8 GPUs, it is possible to explore the dynamics at much higher resolutions, and therefore Reynolds numbers then previously possible.  The calculation shown in Figure \ref{fig:Diffusion_app} using a resolution of $8192^2$ and reaches a magnetic Reynolds number $Re_M = 10^6$, with a magnetic Prandtl number of 1.  In idealized calculations of isolated current sheets it has been found that the magnetic reconnection rate becomes independent of $Re_M$ when $Re_M \geq 10^5$ \citep{loureiro2012magnetic, uzdensky2010fast}, and reconnection enters the plasmoid-dominated regime.  The zoom-in panel on the left of Figure \ref{fig:Diffusion_app} shows that after 10 orbits a large number of plasmoids have formed, although most of the current sheets in the flow are not associated with reconnection.  Currently this calculation is being repeated in full 3D at comparable resolutions and Reynolds numbers to investigate magnetic reconnection and turbulence driven by the MRI at high $Re_M$.  Such calculations are only possible with codes that can exploit exascale resources.

\subsection{Two-Fluid Magnetohydrodynamics}

\begin{figure*}[htb]
    \centering
    \includegraphics[width=3.0in]{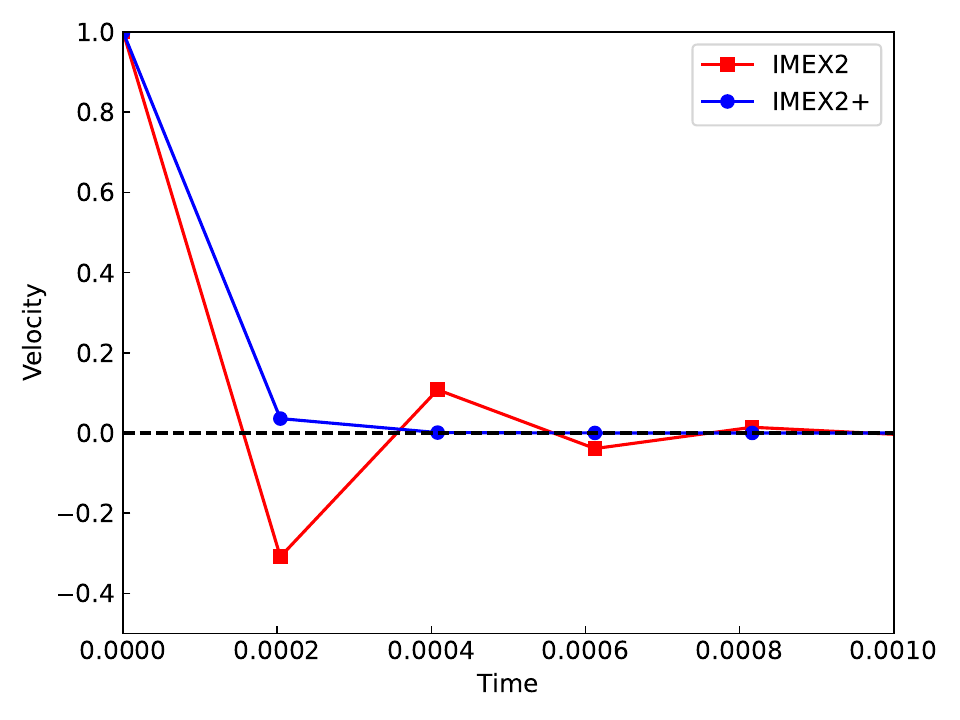}
    \includegraphics[width=3.0in]{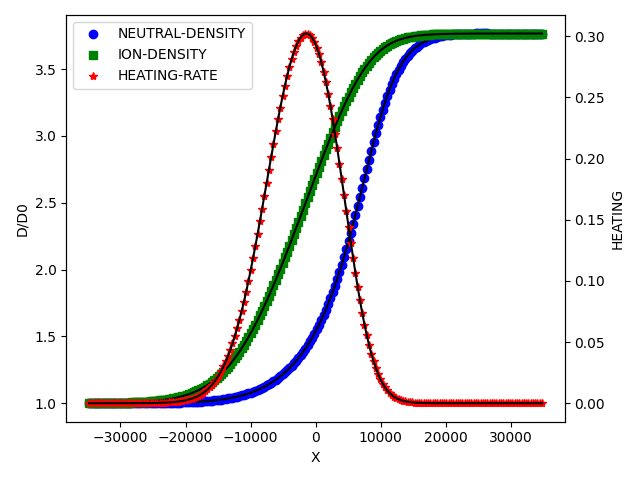}
    \caption{(Left) Decay of the ion velocity using second-order IMEX integrators when the drag source term is very stiff. (Right) Profiles of the ion and neutral densities, and
    per-volume heating rate in a C-type shock.  Points are shown after 10 flow crossing times, while the solid lines are the initial conditions.}
    \label{fig:twofluid_test}
\end{figure*}

As discussed in \S2, physics modules in \ak{} are included as independent objects in the {\tt MeshBlockPack}, each with its own task list.  This enables combining separate modules in the same calculation (for example multiple fluids, or a single fluid and the particle module described in \S4).  As a demonstration, we consider two-fluid MHD, in which both a hydrodynamic and an MHD module are included, with the momentum of the two fluids coupled through a collisional drag term
\begin{equation}
    {\bf S} = \alpha \rho_i \rho_n ({\bf v}_i - {\bf v}_n)
\end{equation}
where $(\rho_i, {\bf v}_i)$ and $(\rho_n, {\bf v}_n)$ are the mass density and velocity of the ions (MHD fluid) and neutrals (hydrodynamic fluid) respectively.  This term is added to (subtracted from) the neutral (ion) momentum equations.  Source terms to account for ionization and recombination, as well collisional drag heating, can also be included, however for simplicity we shall consider only non-reacting and isothermal fluids here.

Since the collisional drag term may be become very large, the two-fluid equations of motion may be stiff, and the source term must be time-integrated implicitly. The system is therefore well suited for IMEX integrators, and in \ak{} we have implemented both second- and third-order IMEX integrators of \cite{PareschiRusso}, as well as the positivity preserving method described by \cite{krapp2024ApJS..271....7K}.  The left panel of Figure \ref{fig:twofluid_test} shows
a comparison of the second-order IMEX integrators using a 1D relaxation test. Two oppositely directed, equal density flows are initialized with $\rho_i = \rho_n = 1$ and $v_i = 1$ and $v_n = -1$, respectively. The initial plasma $\beta$ is 1. The drag coefficient between ions and neutrals is set to $\alpha = 6\times 10^3$.  The two flows should rapidly come to equilibrium at rest.  Data is output at every time step to follow this evolution.  Note the IMEX2 integrator of \cite{PareschiRusso}
results in overshooting followed by damped oscillations toward the solution, whereas that of \cite{krapp2024ApJS..271....7K} smoothly converges to the correct result.  Such overshoot can be very problematic when stiff source terms are added to the energy equation.

The right panel of Figure \ref{fig:twofluid_test} shows the profile of the steady-state C-shock computed by \ak{}.  The initial conditions are computed by solving the ODEs resulting from the assumption of time-independent equations of motion. This solution is mapped to a grid of 200 cells and then evolved with \ak{} for 10 flow times across the domain on a single CPU core. The upstream initial conditions in the shock are $\rho_i = 10^{-3}$, $ \rho_n = 1$, $v_i = v_n = 30$, with a transverse magnetic field $B_y/\sqrt{4 \pi} = 10$, and collisional drag coeffcient $\alpha=1$.  The upstream boundary conditions are held at these values, and the downstream boundary condition is outflow. The isothermal sound speed $C=1$ in both fluids, giving an upstream Mach number of 30, and Alfv\'{e}n Mach number in the neutrals of 3.  The characteristic width of the shock is the neutral-ion coupling length $V_{A,n}/(\alpha \rho_i) = 10^4$. The figure compares the final solution (points) with the initial conditions (solid line).  The excellent correspondence between the two confirms the module correctly hold a steady C-shock profile.

\begin{figure*}[htb]
    \centering
    \includegraphics[width=\linewidth]{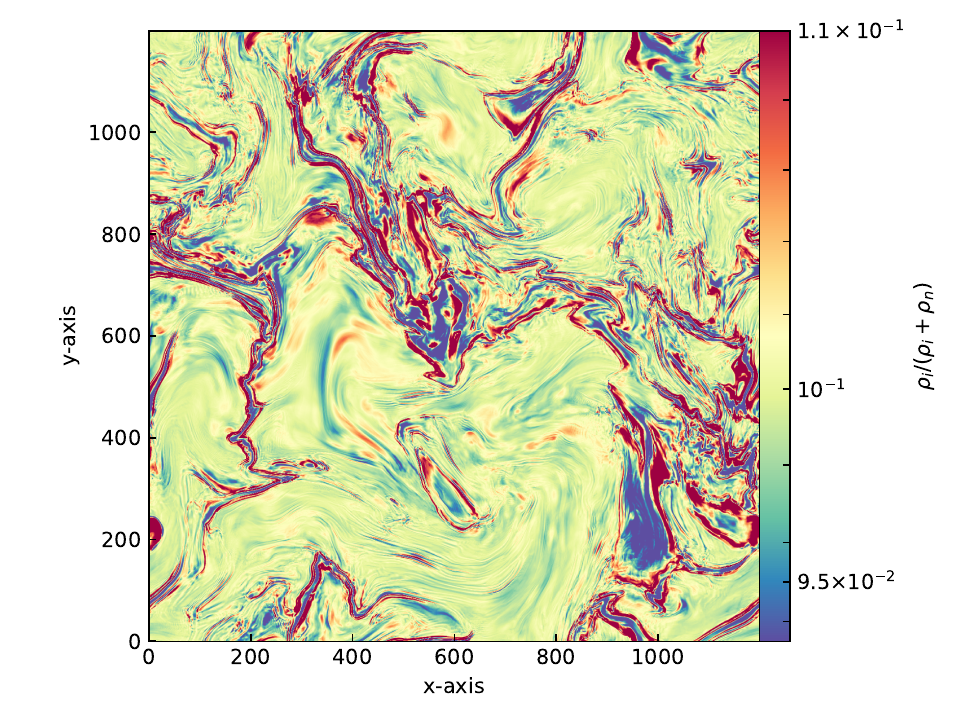}
    \caption{Slice of the ionization fraction in a 3D two-fluid simulation of compressible turbulence using a uniform grid of $1200^3$ cells.}
    \label{fig:Two-fluid_app}
\end{figure*}

\ak{} has already been used for new studies of isothermal two-fluid MHD turbulence in the ISM \citep{hu2024MNRAS.527.3945H}. Figure \ref{fig:Two-fluid_app} shows typical results for the ionization fraction $f_i = \rho_i/(\rho_n + \rho_i)$ in a calculation using periodic boundary conditions, the IMEX3 time integration algorithm, and the PPM4 spatial reconstruction method.  Initially, both the magnetic field and the ion and neutral density fields are uniform, with the magnetic field aligned along the $z-$axis. The initial ionization fraction is set to 0.1, and the drag coefficient between ions and neutrals is $10^6$. The simulation box is divided into $1200^3$ uniform cells.  Turbulence is driven solenoidally at wavenumbers $k = 1 - 2$ to achieve a Kolmogorov-like power spectrum. Snapshots were taken after the turbulence reached statistical saturation and the spectral slope was stable. The simulation ultimately achieves a sonic Mach number of $M_s \approx 10$ and an Alfv\'{e}nic Mach number of $M_A \approx 1$.  The image demonstrates that $f_i$ can show complex fluctuations in two-fluid turbulence.  An important result \citep{hu2024MNRAS.527.3945H} is that the fluctuations in $f_i$
can be of very large amplitude in some cases, although for the very strong coupling used in the calculation shown in Figure \ref{fig:Two-fluid_app}, the fluctuations are small.

\subsection{Special relativistic hydrodynamics and MHD}

In special relativistic MHD (SRMHD), assuming a unit system where $c = 1$, the conserved variables and their fluxes are
\begin{align}
{\bf U} &= \begin{bmatrix}
           D \\
           \mathbf{M} \\
           E - D 
         \end{bmatrix},
\hspace*{0.1cm}{\bf F} = \begin{bmatrix}
           D {\bf v} \\
           \mathbf{S} \\
           \mathbf{M} - D \mathbf{v}
         \end{bmatrix},
\label{eq:srmhd_vars}
\end{align}
where 
\begin{align}
D &= \Gamma \rho, \\
E &= \Gamma^2 w - \Gamma^2 \left(\mathbf{v \cdot B} \right)^2 - \left( p_g + p_m \right), \\
\mathbf{M} &= (E + p_g + p_m) \mathbf{v} - \left( \mathbf{v} \cdot \mathbf{B} \right) \mathbf{B}.
\end{align}
The stress tensor is
\begin{align}
\nonumber
    \mathbf{S} = \Gamma^2 w \mathbf{v} \mathbf{v} &- \frac{1}{\Gamma^2} \mathbf{B} \mathbf{B} - \left(\mathbf{v} \cdot \mathbf{B} \right) \left(\mathbf{v} \mathbf{B} + \mathbf{B} \mathbf{v} \right) 
    \\ 
    &- \Gamma^2 \left(\mathbf{v} \cdot \mathbf{B} \right)^2 \mathbf{v} \mathbf{v} + \left(p_g + p_m \right) \mathbf{I},
\end{align}
the total enthalpy is 
\begin{equation}
    w = \rho + \frac{\gamma}{\gamma - 1} p_g + 2 p_m,
\label{eq:rel_enthalpy}
\end{equation}
and $\Gamma$ is the Lorentz factor \citep{WhiteStone2016,Stone2020ApJS..249....4S}.
The gas pressure $p_g$ is related to the gas internal energy $u_g$ via a $\gamma$-law equation of state $p_g = (\gamma - 1) u_g$ and the magnetic pressure is
\begin{equation}
    p_m = \frac{1}{2} \left( \frac{1}{\Gamma^2} \mathbf{B}^2 + \left( \mathbf{v} \cdot \mathbf{B} \right)^2 \right).
\end{equation}
We subtract the conserved density $D$ from the total energy $E$ in constructing our total energy conserved variable, in analogy to \texttt{HARM}'s treatment of the total energy equation \citep{2003ApJ...589..444G}; this choice makes integrations more accurate in regimes where the magnetic and internal energy densities are much smaller than the rest mass density.  

Special relativistic MHD in \texttt{AthenaK} invokes a primitive state vector $\mathbf{W} = \left[\rho, u^i, u_g \right]$, where $u^i$ are the spatial components of the coordinate 4-velocity $\mathbf{u}$, related to the three velocity via $\mathbf{u} = \Gamma \mathbf{v}$.  The conserved state vector $\mathbf{U}$ can be analytically constructed from the primitive state vector $\mathbf{W}$, however, the inverse cannot.  Here, in a departure from \texttt{Athena++}, we invoke the iterative conserved to primitive (C2P) solver of \cite{2021PhRvD.103b3018K}.

\begin{figure*}[htb]
    \centering
    \includegraphics[width=\linewidth]{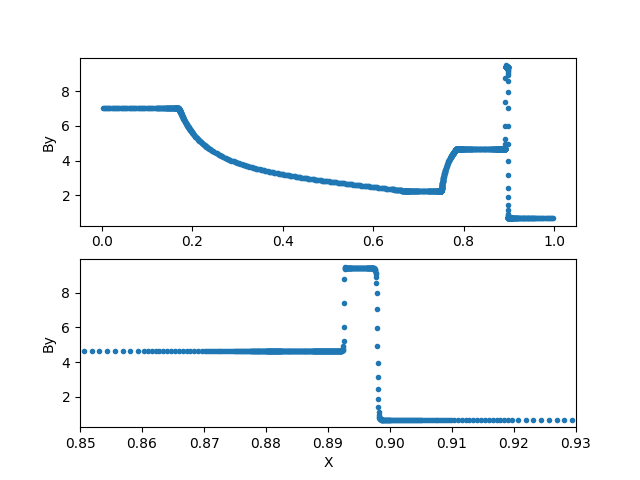}
    \caption{Transverse component of the magnetic field in the SRMHD shock tube test. Zoom-in shown in the bottom panel demonstrates how AMR can capture the very thin shell formed in this test.}
    \label{fig:srmhd-shocktube}
\end{figure*}

Figure \ref{fig:srmhd-shocktube} shows the results for a relativistic MHD blast
wave test given in \S4.3.1 of S20 as proposed by \cite{mignone2012ApJS..198....7M}.  The test uses AMR to
capture the very thin shell formed by the relativistic dynamics.  The initial conditions, grid, and AMR refinement criteria used in this test are identical to those described in S20, although here we use WENOZ reconstruction and the HLLE Riemann solver. The results are shown at a time $t=0.4$ for direct comparison to Figure 37 in S08.
Note that the shell is well resolved with AMR.

\begin{figure*}[htb]
    \centering
    \includegraphics[width=\linewidth]{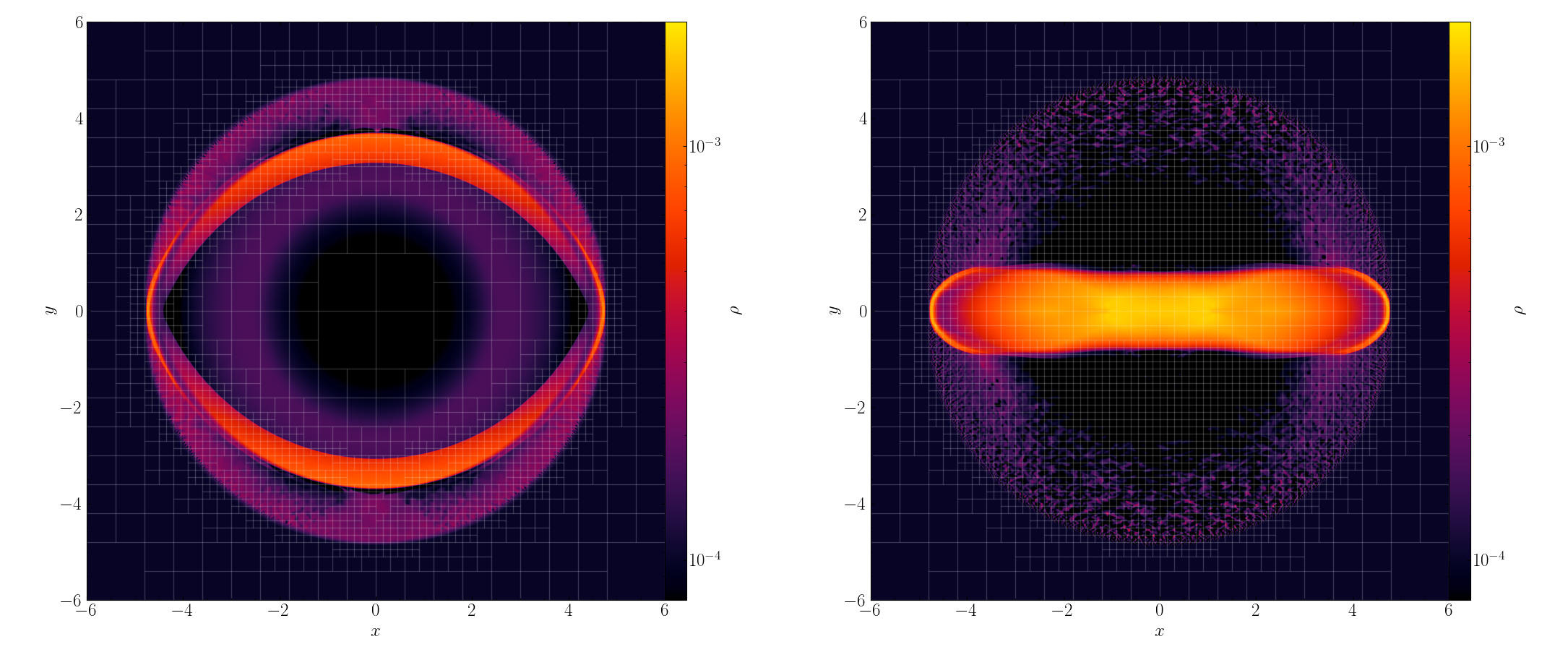}
    \caption{Logarithmic density $\rho$ in the weak field ($B^x = 1/10$, \textit{left}) and strong-field ($B^x = 1$, \textit{right}) \cite{1999MNRAS.303..343K} cylindrical explosion tests, run on an AMR grid with pressure-based refinement.  White lines denote 20 $\times$ 20 \texttt{MeshBlock} boundaries.}
    \label{fig:srmhd_test2}
\end{figure*}

\cite{1999MNRAS.303..343K} devised a series of challenging SRMHD cylindrical blast wave test problems (invoking high $\sigma = 2 p_m / \rho$ ambient backgrounds), where the relative ``difficulty'' of the problem is associated with the choice of initial (uniform) magnetic field strength $B^x$.  Although mostly qualitative, the cylindrical explosion tests certainly challenge the robustness strategies of the SRMHD algorithm (including the C2P solver); many finite volume SRMHD codes require stringent flooring, a modification to the background state \citep[e.g.,][]{2005A&A...436..503L, 2007A&A...473...11D, Beckwith2011}, and/or an artificial viscosity \citep[e.g.,][]{1999MNRAS.303..343K} to evolve the problem.  We find that first-order flux corrections (FOFC) enable us to integrate the problem robustly.  Figure \ref{fig:srmhd_test2} shows \texttt{AthenaK} solutions for the cylindrical explosion tests, now supplemented with pressure-based AMR refinement.  The blast is prescribed via an over-dense ($\rho = 10^{-2}$), over-pressurized ($p_g = 1.0$) region (within cylindrical radius $r \leq 1$), amidst an ambient background of gas density $\rho = 10^{-4}$ and pressure $p_g = 3 \times 10^{-5}$. An exponential ramp \citep[see e.g.,][]{2014CQGra..31a5005M} is used to smoothly bridge the two regions.  Figure \ref{fig:srmhd_test2} (\textit{left}) and (\textit{right}) correspond to initial field strengths of $B^x = 0.1$ and $B^x = 1$, respectively.

\begin{figure}[htb]
    \centering
    \includegraphics[width=\linewidth]{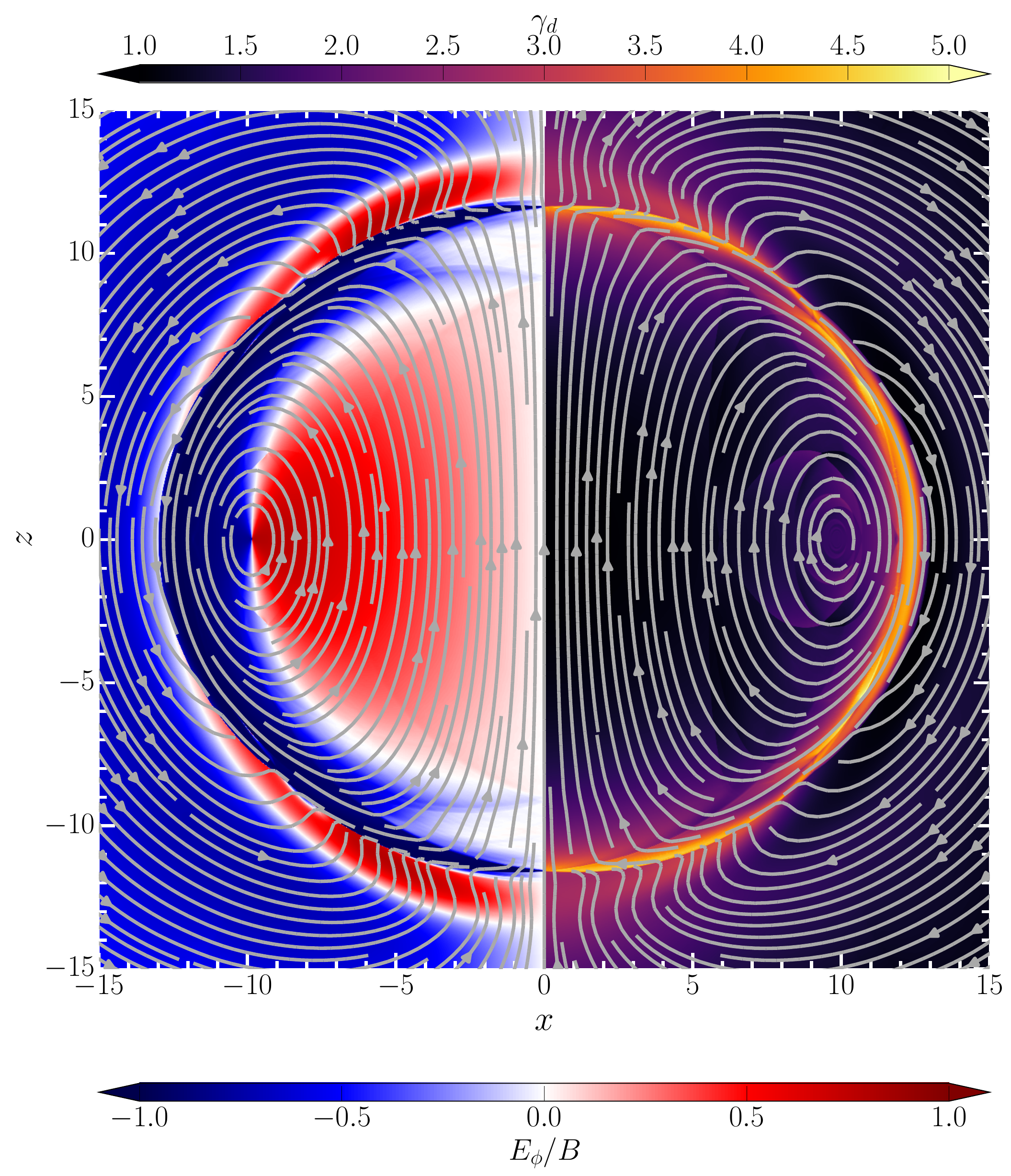}
    \caption{Formation of a monster shock inside a dipole magnetosphere (shown with grey lines). Shown are the drift Lorentz factor $\gamma_d$ and the out-of-plane electric field component $E_\phi$, normalized to the magnetic field $B$.}
    \label{fig:monster}
\end{figure}

As an additional application for the SRMHD module, we study the formation
of a strong shock inside a magnetar magnetosphere. It has recently been
shown that magnetosonic fast waves inside MHD dipole magnetospheres can steepen into {\it monster shocks} \citep{2023ApJ...959...34B}. This can have implications for novel
gamma-ray burst mechanisms from collapsing magnetars \citep{2024arXiv240401456M}, as well as 
for FRB mechanisms inside magnetar magnetospheres \citep{2024arXiv240715076V}.  Within the geometry of the dipole
magnetosphere, this will happen when electric field toroidal component $E_\phi$ approaches the background magnetic field strength $B$.
In a strongly magnetized magnetosphere with magnetization $\sigma =
b^2/\rho \gg 1$ (where $b$ is the magnetic field strength in the comoving
frame), the limiting value of the magnetosonic speed $c_s$ will be set by
$c_s^2 \approx 1 - 1 /\sigma$ \citep{2023ApJ...959...34B}. The fact that $\sigma$ is finite is
a requirement of the shock, as shock formation requires self-regulation of
the shock by means of plasma heating. This makes simulations of monster
shock formation extremely challenging, as it combines high Lorentz factors
and magnetizations. We show the formation of a monster shock for an
idealized two-dimensional scenario in Figure \ref{fig:monster} 
using an $8192^2$ mesh on a square domain with edges spanning $[-20, 20]$. 
For this simulation, we adopt a $\gamma=4/3$ ideal
fluid equation of state, setup a dipole magnetic field (as shown in Fig.
\ref{fig:monster}), and initialize the temperature and density profiles by
demanding that the magnetization parameters $\beta= 0.1$ and $\sigma = 40$
take uniform values.

In Figure \ref{fig:monster}, we show the early onset of shock formation.
The shock forms when $E_\phi/ B \rightarrow 1$. The shock resembles the
simulation shown in \citet{2024arXiv240401456M}.
A full analysis of shock propagation, as well as on the dynamics of the
shock will be provided in a separate study (Smith et al., \textit{in prep}).

\subsection{General relativistic hydrodynamics and MHD}

\begin{figure*}[htb]
    \centering
    \includegraphics[width=\linewidth]{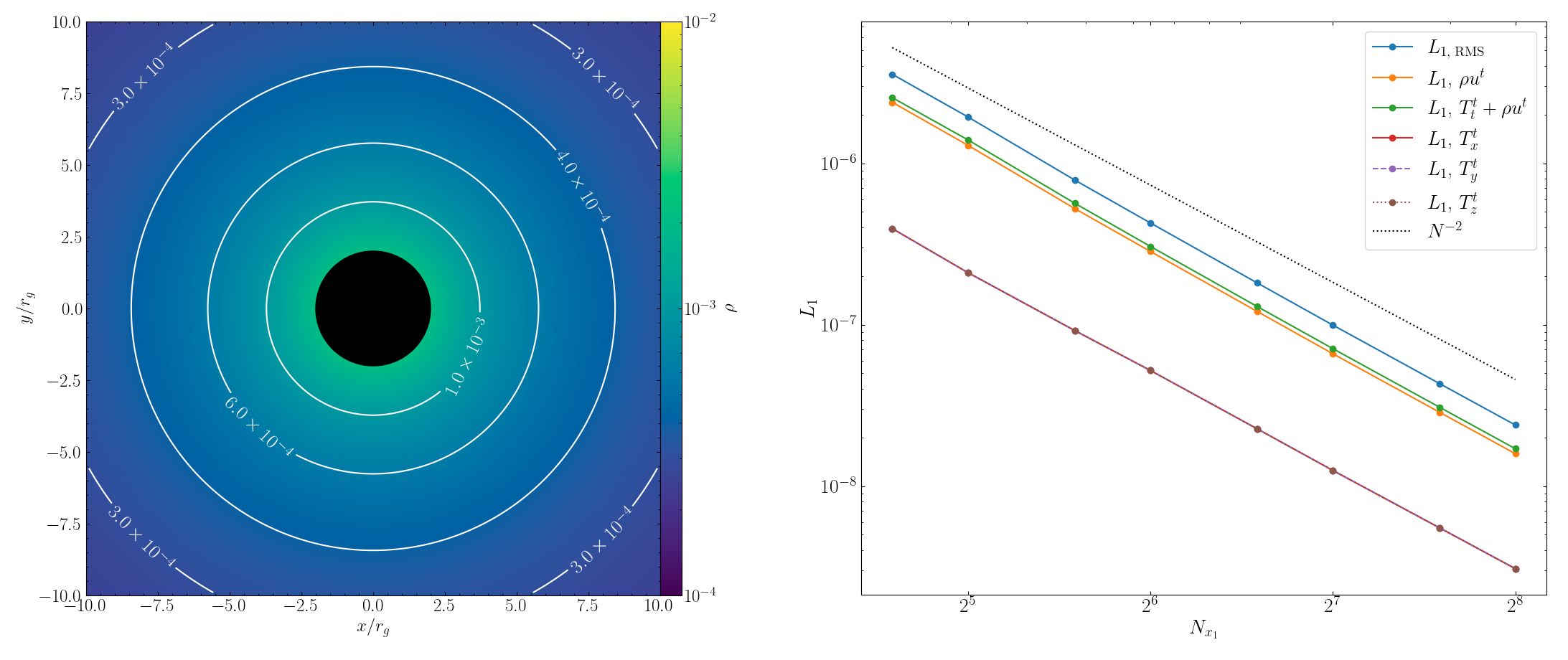}
    \caption{(Left): Logarithmic midplane density $\rho$ (i.e., $z=0$ slice) and contours from a spherically-symmetric Bondi accretion test problem; (Right): Convergence with numerical resolution for conserved variables $\rho u^t, T^t_\mu$; $\propto N^{-2}$ convergence is shown in dotted black.}
    \label{fig:GRMHD_test1}
\end{figure*}

\begin{figure*}[htb]
    \centering
    \includegraphics[width=\linewidth]{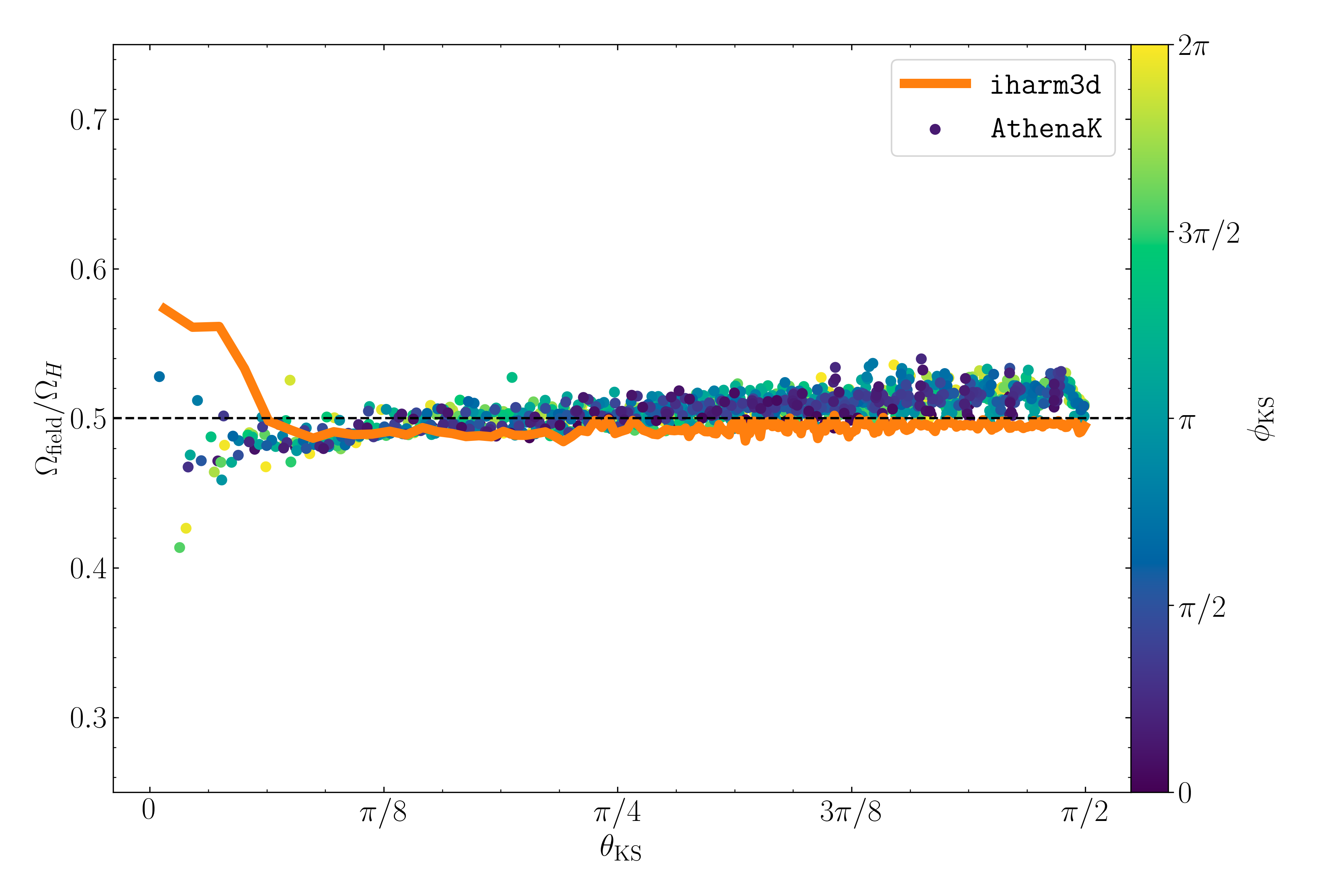}
    \caption{Field rotation rate $\Omega_\mathrm{field}$ to horizon angular frequency $\Omega_H$ ratio measured at the horizon in a BZ monopole test problem (spin $a=1/2$).  Results from a 2D curvilinear \texttt{iharm3d} simulation are shown in orange.  Results from a 3D CKS \texttt{AthenaK} simulation are shown at various $\theta$, $\phi$ sample points.  The expected solution $\Omega_\mathrm{field} / \Omega_{H} \simeq 1/2$ is shown in dashed black.}
    \label{fig:GRMHD_test2}
\end{figure*}

\begin{figure*}[htb]
    \centering
    \includegraphics[width=\linewidth]{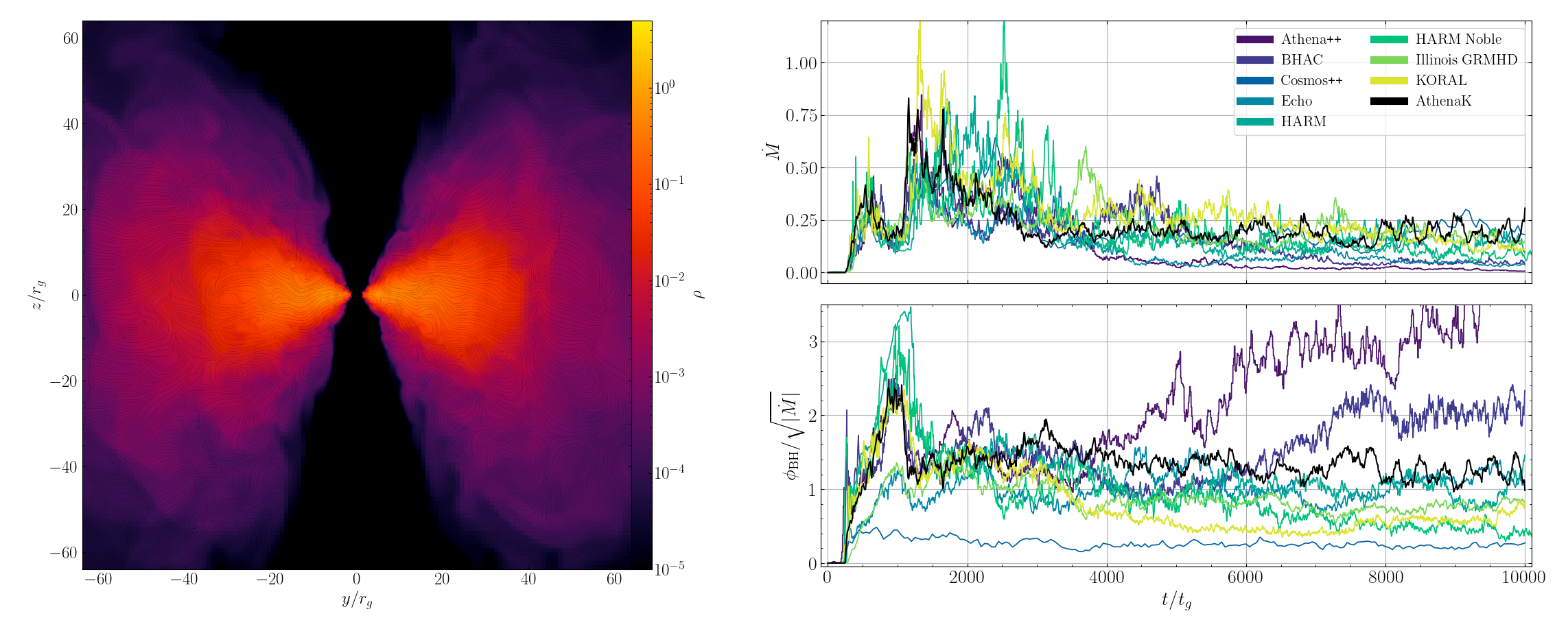}
    \caption{(Left): Snapshot of logarithmic density $\rho$ in a poloidal slice through a SANE Fishbone-Moncrief torus about an $a = 15/16$ black hole evolved to $t \simeq 4000 \; t_g$; streamlines trace spatial components of the coordinate frame, contravariant 4-velocity (Right): Mass accretion rates $\dot{M}$ and (normalized) magnetic flux $\phi_\mathrm{BH} / \sqrt{\lvert \dot{M} \rvert}$ from the Porth et al. (2019), EHT SANE code comparison project are presented with \texttt{AthenaK} solutions overplotted in black.}
    \label{fig:GRMHD_app}
\end{figure*}

\texttt{AthenaK}'s general relativistic MHD (GRMHD) module evolves (assuming a unit system where $GM = c = 1$)
\begin{equation}
    \partial_t \mathbf{U} + \partial_j \mathbf{F} = \mathbf{S},
\label{eq:grmhd_einstein}
\end{equation}
where the conserved variables, their fluxes, and source terms \textit{associated with the connection} are
\begin{align}
\nonumber
{\bf U} = \sqrt{-g} \begin{bmatrix}
           \rho u^t \\
           T^t_i \\
           T^t_t + \rho u^t
         \end{bmatrix},
\hspace*{0.1cm}{\bf F} &= \sqrt{-g} \begin{bmatrix}
           \rho u^j \\
           T^j_i \\
           T^j_t + \rho u^j
         \end{bmatrix},
         \\
{\bf S} &= \sqrt{-g} \begin{bmatrix}
           0 \\
           \dfrac{1}{2} \left( \partial_i g_{\alpha \beta} \right) T^{\alpha \beta} \\
           0
         \end{bmatrix},
\label{eq:grmhd_vars}
\end{align}
respectively, where $g = \det {\mathbf{g}}$ and $\mathbf{g}$ is the metric, the rest-mass density is $\rho$, the coordinate frame 4-velocity is $u^\mu$, the stress-energy tensor is
\begin{equation}
T^\mu_\nu = w u^\mu u_\nu - b^\mu b_\nu + (p_g + p_m) \delta^\mu_\nu,
\end{equation}
the magnetic pressure is derived from the magnetic field $b^\mu$ via
\begin{equation}
    p_m = \frac{1}{2} b_\mu b^{\mu},
\end{equation}
and $w$ is the total enthalpy \citep{2003ApJ...589..444G, WhiteStone2016, Stone2020ApJS..249....4S}.  We again apply the \texttt{HARM}--``trick'' \citep{2003ApJ...589..444G}, wherein we eliminate contributions from the conserved density $\rho u^t$ from the evolved total energy conserved variable.

The GRMHD primitive state vector is $\mathbf{W} = \left [ \rho, \tilde{u}^i, u_g \right]$, where $\tilde{u}^i$ are normal frame, spatial velocity components, related to the coordinate frame velocity via $u^i = \tilde{u}^i - \beta^i \Gamma / \alpha$, where $\alpha$, $\beta^i$, and $\Gamma$ are the lapse, shift, and normal frame Lorentz factor, respectively.  GRMHD shares the same \cite{2021PhRvD.103b3018K} C2P solver as SRMHD, supplemented only with the necessary frame transformations to cast the fluid state to the SR frame.  

The GRMHD induction equation
\begin{equation}
    \partial_t \left( \sqrt{-g} B^i \right) + \partial_j \left( \sqrt{-g} \left(b^i u^j - b^j u^i \right) \right) = 0
\end{equation}
evolves the magnetic field 3-vector components $B^i$ subject to the no-monopole constraint
\begin{equation}
    \partial_j \left( \sqrt{-g} B^j \right) = 0,
\end{equation}
where $B^i$ is related to the contravariant magnetic field $b^\mu$ via
\begin{align}
b^t &= u_i B^i, \\
b^i &= \frac{1}{u^t} \left(B^i + b^t u^i \right).
\label{eq:dual_faraday}
\end{align}
\texttt{AthenaK}'s GRMHD module is Cartesian only, invoking either the Cartesian Kerr-Schild \citep[CKS;][]{2007arXiv0706.0622V} or Minkowski metric, where both share $\sqrt{-g} = 1$.  Following \cite{2003ApJ...589..444G}, Equations (\ref{eq:grmhd_einstein}, \ref{eq:grmhd_vars}) are presented in \textit{free index down} form, hence giving rise to a conserved energy current.  Because \ak{}'s GRMHD module operates in Cartesian only, our algorithm does not conserve angular momentum to machine precision as \texttt{HARM}.  

When operating in CKS, an ``excision'' or ``inner boundary'' about the black hole is exercised \citep[c.f.,][]{2021MNRAS.504.6076R, 2019A&A...629A..61O}; \texttt{AthenaK} defaults to setting $\rho = \rho_\mathrm{excision}$, $p_g = p_{g, \mathrm{excision}}$, and $\tilde{u}^i = 0$ for $r \leq \frac{1}{2} \left(r_{h, \mathrm{outer}} + r_{h, \mathrm{inner}} \right)$, where $r_{h, \mathrm{outer}}$ and $r_{h, \mathrm{inner}}$ are the outer and inner horizon radii, respectively, evaluated in Kerr-Schild coordinates (hereafter, $r, \theta, \phi$).

Spherically-symmetric, Bondi accretion onto a Schwarzschild black hole is a common test problem for GR hydrodynamics codes;  typically this test is performed in (spherical) Kerr-Schild coordinates (or modified variants thereof) \textit{in 1D or 2D} \citep[e.g.,][to name a few]{2003ApJ...589..444G, WhiteStone2016, 2017ComAC...4....1P, 2021JOSS....6.3336P, 2024MNRAS.532.3198C}.  CKS coordinates require \ak{} to invoke a full 3D discretization for this problem.   Our initial conditions (with black hole spin $a = 0$) are those prescribed by \cite{1984ApJ...277..296H} and \cite{WhiteStone2016}, invoking a sonic point radius $r_c = 8M$, a polytropic index equal to the adiabatic index (i.e., isentropic) of $\gamma = 4/3$, and an adiabat $K = p_g / \rho^\gamma = 1$.   Our simulation domain is cubic with edge lengths of $20M$.  Outer boundary conditions are prescribed to the analytic solution.  Our excision sets the primitive state vector to floors interior to the horizon ($r_h \leq 2M$), as previously detailed.   For $r_h \leq r \leq 3M$, we reset the solution to analytic values after every integration stage --- in this manner, noise introduced at the excision boundary does not enter our convergence analysis for all resolutions surveyed.  Figure \ref{fig:GRMHD_test1} (\textit{left}) plots the rest mass density in an equatorial slice (i.e., $z=0$) through the simulation domain after evolving the solution for $t = 100M$.  We overplot contours to highlight \ak{}'s ability to preserve the spherically symmetric nature of the problem, despite operating in CKS coordinates.  More quantitatively, Figure \ref{fig:GRMHD_test1} (\textit{right}) presents a convergence analysis demonstrating second order convergence in $L_1$ errors of the conserved state vector when applying an RK2 temporal integrator, PLM reconstruction, and an HLLE Riemann solver.  

Following \cite{2024MNRAS.532.3198C}, we next apply \ak{} to the so-called Blandford \& Znajek \citep[BZ;][]{1977MNRAS.179..433B} monopole test problem.  A monopole magnetic field $B^r \propto 1/r^2$ (initialized through a toroidal vector potential) threads a spherically symmetric medium prescribed via a density power-law
\begin{equation}
\rho(r) = \frac{\left( B^r \right) ^2}{\sigma_c} \left( \frac{r}{r_c} \right)^{-n}
\end{equation}
about a spin $a = 1/2$ black hole.  We parameterize the gas pressure via $p = \rho T$ with $T = \gamma - 1$.  We select $r_c = 10 \; r_{h, \mathrm{outer}}$ and $n = -1$, which still corresponds to a density falloff with radius given the $\propto r^{-4}$ dependence on $\left( B^r \right) ^2$.   $\sigma_c$ determines the ratio of the magnetic pressure to the gas density; we select $\sigma_c = 100$, which challenges \ak{}'s robustness strategies and primitive recovery scheme.   At its heart, the BZ monopole problem tests our algorithm's ability to recover the expected field rotation rate
\begin{equation}
    \Omega_\mathrm{field} = \frac{b^r u^\phi - b^\phi u^r }{b^r u^t - b^t u^r}
\label{eq:field_rot_rate}
\end{equation}
about the horizon, amidst a challenging fluid state subject to flooring, first order flux corrections, and/or C2P failure mitigation strategies.  A $256 \times 256 \times 128$ grid resolves a domain with boundaries $x \in [-10M, 10M]$, $y \in [-10M, 10M]$, $z \in [0, 10M]$. Outflow boundary conditions are applied everywhere except the $z=0$ plane, where we instead enforce reflecting boundary conditions.  After evolving for $t = 10M$, we measure the field rotation rate $\Omega_\mathrm{field}$ at the outer horizon by (1) interpolating the CKS primitive state vector to a surface $r = r_{h, \mathrm{outer}}$, (2) converting to Kerr-Schild coordinates, then (3) evaluating Equation \ref{eq:field_rot_rate} at $\theta, \phi$ sample points.  In Figure \ref{fig:GRMHD_test2}, we plot $\Omega_\mathrm{field} / \Omega_h$ where $\Omega_h$ is the rotation rate of the horizon.  We overplot both the expected solution $\Omega_\mathrm{field} / \Omega_h = 1/2$ and a solution generated with \texttt{iharm3D} \citep{2021JOSS....6.3336P}.
 
Finally, we report \ak{} accretion rate histories for the \texttt{SANE} EHT code comparison run \citep{2019ApJS..243...26P}.  A Fishbone-Moncfrief \citep[FM;][]{1976ApJ...207..962F} torus ($r_\mathrm{in} = 6M$; $r_\mathrm{max} = 12M$) is initialized about a spin $a = 15/16$ black hole.  We match the code comparison project by (1) invoking a weak poloidal field loop initialized via a toroidal vector potential (with a strength parameterized by the global maxima of gas and magnetic pressures), (2) an adiabatic index $\gamma = 4/3$, (3) a 4\% perturbation added to the gas pressure, and (4) evolving to $t = 10^4 M$.  In Figure \ref{fig:GRMHD_app} (\textit{left}), we plot a snapshot of the simulation at $t=4000M$; the poloidal slice of the (logarithmic) gas density field is decorated with streamlines corresponding to spatial components of the coordinate velocity.  Figure \ref{fig:GRMHD_app} (\textit{right}) plots the \ak{} mass accretion rate $\dot{M}$ and (normalized) magnetic flux at the horizon $\phi_\mathrm{BH}$ alongside the $128^3$ datasets reported in \cite{2019ApJS..243...26P} for a suite of GRMHD codes.  Many of the other codes surveyed in Figure \ref{fig:GRMHD_app} use logarithmic gridding to focus resolution near the horizon making a $1:1$ comparison with our CKS grid less straightforward.

\subsection{General relativistic radiation MHD}

\begin{figure*}[htb]
    \centering
    \includegraphics[width=\linewidth]{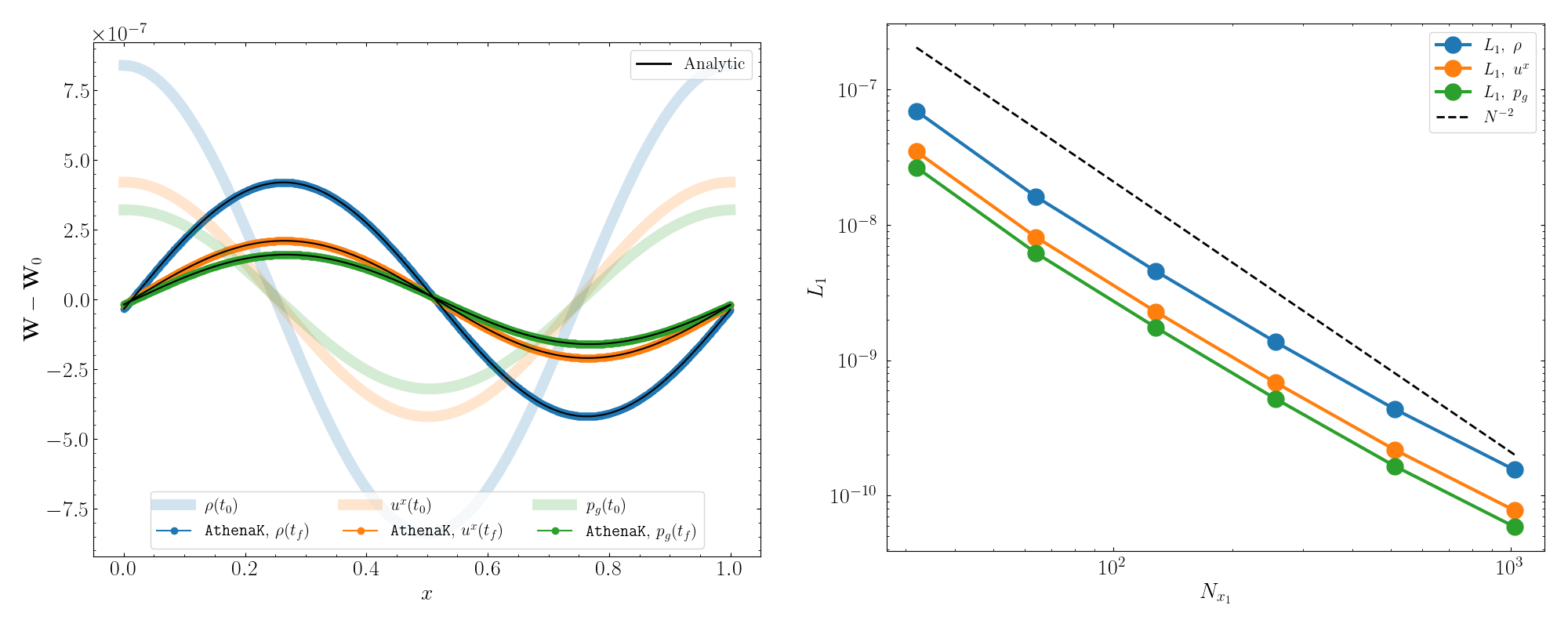}
    \caption{(Left): \texttt{AthenaK} primitive solutions $\mathbf{W}$ for a radiation-modified linear wave test problem at initial time $t_0$ and final time $t_f$; fluid frame density $\rho$, coordinate frame 4-velocity spatial component $u^x$, and fluid frame gas pressure $p_g$ are shown in blue, orange, and green respectively;  analytic solutions are presented in black. (Right) Convergence with numerical resolution for primitives $\mathbf{W}$; $\propto N^{-2}$ convergence is shown in black.}
    \label{fig:GRrad_test1}
\end{figure*}

\begin{figure*}[htb]
    \centering
    \includegraphics[width=\linewidth]{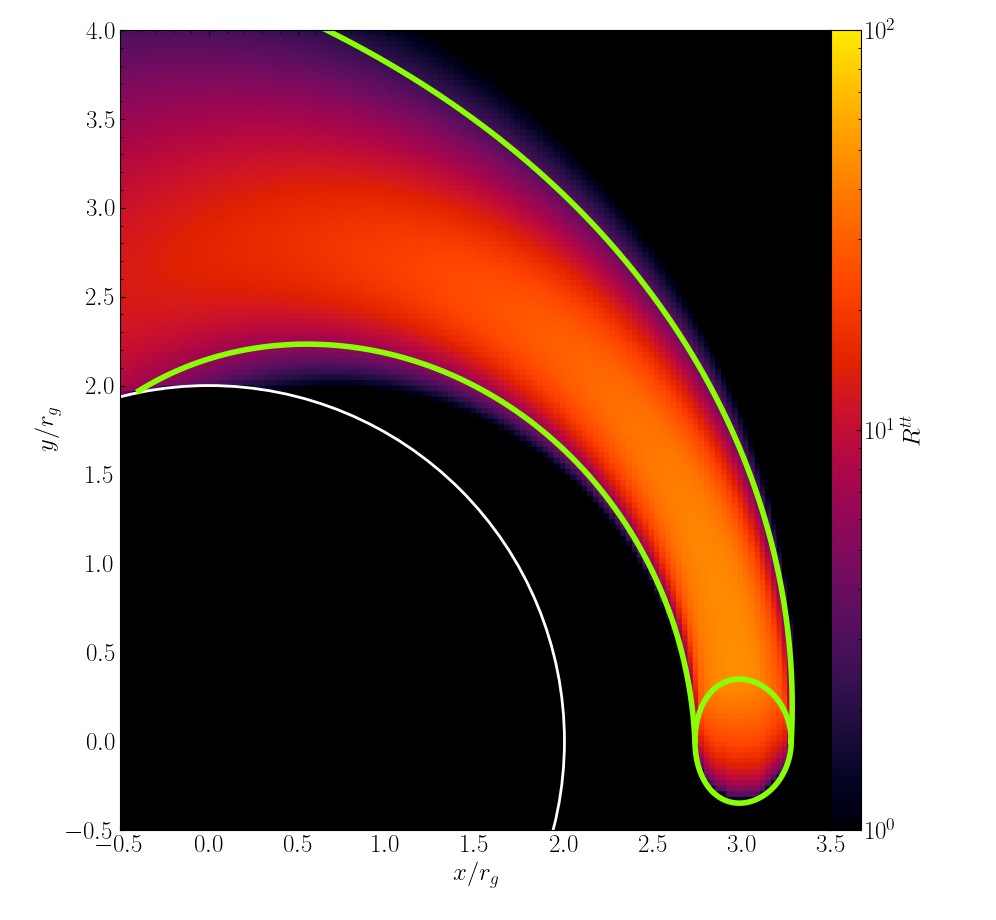}
    \caption{Beam propagation in the Schwarzschild metric.  Color shows logaramithic, coordinate frame radiation energy density $R^{tt}$ for an equatorial beam injected along the photon orbit.  Lime green tracks geodesics along beam boundaries; white traces the event horizon.}
    \label{fig:GRrad_test2}
\end{figure*}

\begin{figure*}[htb]
    \centering
    \includegraphics[width=\linewidth]{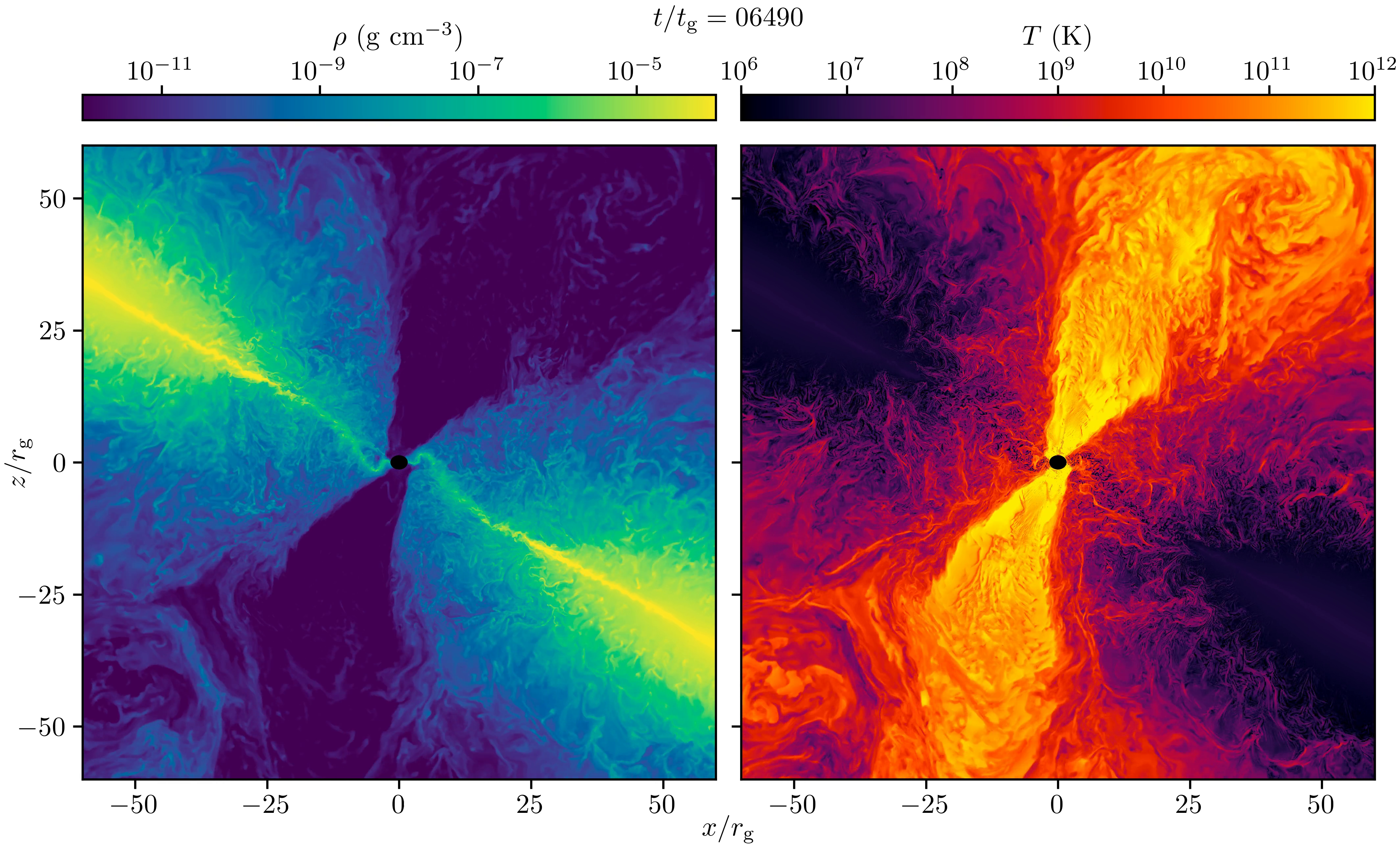}
    \caption{Logarithmic fluid frame mass density $\rho$ (left) and gas temperature $T$ (right) in a poloidal slice through a tilted ($45^\circ$), sub-Eddington accretion disk about an $a = 15/16$ black hole.}
    \label{fig:GRrad_app}
\end{figure*}

The \textit{gray} (i.e., frequency integrated) photon transport equation in general relativity is 
\begin{align}
\nonumber
    \partial_t \left[ n^t n_t I \right] &+ \frac{1}{\sqrt{-g}} \partial_i \left[ \sqrt{-g} n^i n_t I \right] \\
\nonumber
    &+
    \frac{1}{\sin \zeta} \partial_\zeta \left[ \sin \zeta \left( n^\zeta n_t I \right) \right] 
    \\
    &+
    \partial_\psi \left[ n^\psi n_t I \right] = n_t (j - \alpha I),
\label{eq:gr_transport}
\end{align}
where $I$, $j$, and $\alpha$ are the frequency integrated specific intensity, emissivity, and absorptivity, respectively \citep{2020ApJ...888...94D, 2023ApJ...949..103W}.  The (gray) specific intensity is defined in a \textit{tetrad frame} and is a function of spacetime coordinates $x^\alpha$ and momentum space ``angular'' coordinates $\zeta$ (polar) and $\psi$ (azimuthal).  Components of $\mathbf{n}$ help define directions in spacetime coordinates and momentum space, relating the tetrad frame and coordinate frame.  \ak{} employs a \textit{geodesic grid} \citep{2002CSE.....4e..32R, 2011SJSC...33.2536W, 2013ApJS..205...19F, 2023ApJ...949..103W, 2024arXiv240605126D} to discretize solid angle $\Omega$, tiling the unit sphere into hexagons and pentagons.  The ``level'' of the geodesic grid determines the number of angles discretizing momentum space, $N_\mathrm{angles} = 10 N_\mathrm{level}^2 + 2$.  

The conserved variable for the transport equation is $\mathbf{U} = n^t n_t I$, with a corresponding flux $\mathbf{F}$ having contributions from both a ``spatial'' operator (under $\partial_i$) and ``angular'' operators (under $\partial_\zeta$ and $\partial_\psi$).  Fluxes are computed via simple upwinding.  For the radiation source term $\mathbf{S} = n_t \left(j - \alpha I \right)$, \ak{} decomposes the frequency integrated emissivity and absorptivity into absorption and scattering components, where the former assumes the gray form of Kirchoff's law (hence assuming LTE) and the latter assumes \textit{fluid frame} isotropic, elastic scattering.  Only \textit{constant} or \textit{power-law} (i.e., Kramers--like) specific absorption opacities $\kappa_a$ and/or \textit{constant} specific scattering opacities $\kappa_s$ are currently supported. In an operator split update, a locally-implicit, backward Euler discretization advances the fluid frame specific intensity due to absorption and/or scattering.   Radiation talks to the fluid sector through its moments
\begin{equation}
    \partial_t T^t_\mu = - \partial_t R^t_\mu,
\label{eq:rad_couple}
\end{equation}
where $R^t_\mu = \int n^t n_\mu I d \Omega$ are the coordinate frame radiation moments.
Following the operator split update for the radiation source term, we conservatively update the fluid four-momentum via changes in $R^t_\mu$.

Figure \ref{fig:GRrad_test1} shows \ak{} numerical solutions (\textit{left}) and a convergence analysis (\textit{right}) for a gas-dominated, radiation-modified linear wave test in Minkowski space, as described in \cite{2023ApJ...949..103W}.  We must invoke a single-angle-per-octant angular grid for this test problem so that $R^{\mu \nu} \propto \mathrm{diag} \left[ 1, 1/3, 1/3, 1/3 \right]$ (i.e., Eddington) is exactly recovered (as assumed by the analytics).  Figure \ref{fig:GRrad_test1} (\textit{left}) shows the primitive state vector $\mathbf{W}$ for (1) the initial condition (\textit{translucent}) and at (2) a final time (\textit{opaque}), characterized as the time when the seeded linear perturbation decays by a factor of 2.  Analytic solutions are overplotted in black.  Figure \ref{fig:GRrad_test1} (\textit{right}) shows the corresponding convergence analysis, decomposing the $L_1$ errors into contributions from the mass density $\rho$, coordinate frame 4-velocity spatial component $u^x$, and gas pressure $p_g$.  $\propto N^{-2}$ convergence is overplotted in dashed black, but we remark that the expected convergence rate for this problem is non-trivial.  We apply the RK2 temporal integrator with PLM reconstruction for both the hydrodynamics and radiation sector, corresponding to an overall 2nd order error associated with advection (we note that radiation ``angular'' fluxes are zero for this problem); however, \ak{}'s backward Euler, operator split discretization for the radiation source term limits our algorithm to be only first order accurate in time.  The better-than-first-order convergence in Figure \ref{fig:GRrad_test1} (\textit{right}) is a testament to the dominance of advection errors for this gas-dominated linear wave.   We refer readers to \cite{2023ApJ...949..103W} to see the convergence behavior associated with radiation-modified linear waves in different regimes.  Albeit we here only present solutions in 1D, we have confirmed that \ak{} can recover analytics in 3D when inclining the wavevector to the grid \citep[akin to][]{gs08}, \textit{as long as} the angular grid is equally rotated to enforce Eddington.  

Figure \ref{fig:GRrad_test2} shows \ak{} solutions for vacuum transport in the Schwarzschild metric \citep[c.f.,][]{2013MNRAS.429.3533S, 2020ApJ...900...71A, 2020ApJ...901...96A, 2023ApJ...949..103W}.  Radiation is injected along the photon orbit ($r = 3M$) around a non-spinning black hole, with a beam opening width of $0.7M$ (proper distance) about the injection point and a (half-)opening angle of $10^\circ$.  The angular resolution for this problem is defined in terms of the mean number of angles resolving the solid angle of the beam for illuminated cells, $\mathcal{N}_\mathrm{avg} = 144$.  The spatial resolution is $\Delta x = \Delta y = (1 / 32)M$.  Lime green lines depict the beam injection area and analytic geodesics for the beam inner and outer boundaries.  Numerical diffusion of the beam outside the analytic boundaries are associated with truncation error in the spatial and ``angular'' transport operators. The error is likely dominated by our ``angular'' fluxes, which invoke donor cell reconstruction; spatial fluxes invoke PLM. 

Finally, Figure \ref{fig:GRrad_app} highlights a sample production calculation run on OLCF's Frontier using \ak{}'s radiation GRMHD module. A tilted ($45^\circ$) Chakrabarti torus \citep{1985ApJ...288....1C} is initialized in $\sim$hydrostatic and thermal equilibrium about an $a=15/16$ black hole. 
A weak poloidal loop threads the torus.  An $N_\mathrm{level} = 2$ geodesic grid (42 angles) discretizes momentum space.  A nested SMR grid is used to focus resolution about the horizon.  A density unit is selected such that the absorption and scattering opacities lend the system to a sub-Eddington accretion rate.  At $t \simeq 6500M$, the MRI is well-developed in the tilted accretion disk, and a thin, overdense midplane forms due to strong radiative cooling.  MRI- and radiation-mediated angular momentum transport drive mass accretion in the system; GR effects produce a black-hole-spin-aligned, inner accretion disk, while the outer disk remains misaligned \citep{1975ApJ...195L..65B, 1983MNRAS.202.1181P}.

\section{Particle Module}

In addition to the fluid solvers described above, a module for integrating the motion of particles in various regimes has been implemented in \ak{}.  This module is similar to the particle module implemented in \app{} for evolving dust particles \citep{Bai-particles} and cosmic rays \citep{athena-particle-cr}, which was used as the basis for the {\tt Pegasus} hybrid-PIC code \citep{pegasus}.  However, there are some important differences.

In \ak{} an arbitrary number of \texttt{double} and \texttt{int} data for each particle can stored in 2D {\tt View}s.  Each rank stores the data for all particles located within the {\tt MeshBlock}s in its {\tt MeshBlockPack}. Given the {\tt MeshBlock} \texttt{ID} and coordinates of a particle, it is very easy to calculate the integer indices of the cell containing that particle on the regular Cartesian mesh.  This makes interpolating cell data to the particle positions straightforward.

Communication of particle data between \texttt{MPI} ranks proceeds as follows.  At each time step, a list of all particles that will move beyond the boundaries of
the {\tt MeshBlock}s stored on the rank is created, and using the AMR binary tree, the \texttt{ID} of the new {\tt MeshBlock} the particle will enter is found.  This information is then shared amongst all ranks using a gather-scatter operation. Once all ranks have the list of all particles that need to be communicated and their destination, it is simple to calculate how many particles must be sent to or received from each rank, and to post non-blocking \texttt{MPI} sends and receives for the particle data. Particle data is added and removed from the 2D {\tt View}s, which are resized as necessary.  The use of gather-scatter operations is a significant design difference from the manner in which particles are communicated in \app{}, and helps to reduce the total number of messages communicated.

Particle data can be output in a variety of formats, including the \texttt{VTK} single-point type which can be read by many standard visualization tools.  In addition, selected particles can be labelled as tracers, with the mesh data at the location of the particle output to a custom file format at a user-defined interval.  This is very useful for tracing the time-history of the flow sampled by particles.

Other than the data, communication, and I/O infrastructure, the only other ingredients to the particle module are a pusher to integrate the motion of the particles, and a method to interpolate/deposit grid data to particles/cells.  These depend on the types of particles used, and are described in the subsections below.

\subsection{Lagrangian tracer particles}

Rather than updating particle positions by interpolating the fluid velocity to the location of the particle, \citet{2013MNRAS.435.1426G} describes an alternative method to probabilistically move particles according to mass fluxes computed during the fluid evolution step. In this ``Lagrangian Monte Carlo'' approach, particles are localized only to an individual zone rather than to a precise position within the domain and are probabilistically moved between zones according to the finite-volume conserved mass fluxes through faces between neighboring zones. This method forces the particles to follow the fluid flow (up to statistical noise and diffusion), which lies in contrast to velocity update methods, which have been shown to yield to misrepresentative mass distributions and measures of the fluid power spectrum \citep{2013MNRAS.435.1426G}.

Particle positions are updated at each time step. In each cell, the outgoing mass fluxes through cell faces are taken from the fluid calculation and normalized by the conserved mass in the source cell. A random number is then generated for each particle within the cell and used to select whether the particle should remain in the cell or be moved to one of the neighbors according to the normalized outgoing fluxes. Particles are moved between \texttt{MeshBlock}s as described above; however, an extra post-transfer step is required when a particle moves into a \texttt{MeshBlock} at a different refinement level. If the refinement level of the new \texttt{MeshBlock} is less than the original refinement level, the particle is simply moved into the center of the corresponding cell. When the new \texttt{MeshBlock} refinement level is greater than the original refinement level, a random number is generated and used to select which of the four refined cells the particles should be moved into according to the relative ratio of (corrected) mass fluxes into each of the refined cells.

\begin{figure*}[htb]
    \centering
    \includegraphics[width=\linewidth]{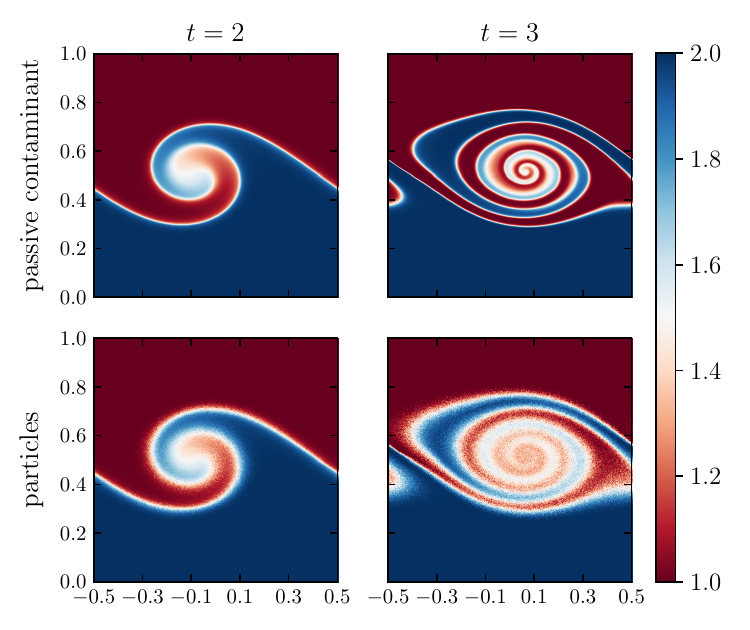}
    \caption{Passive contaminant (top row) and Lagrangian tracer particles (bottom row) in the \cite{2016MNRAS.455.4274L} KHI test.}
    \label{fig:particle-KH}
\end{figure*}

Figure \ref{fig:particle-KH} compares the distribution of a passive contaminant advected with the fluid flow to a set of $10^7$ Lagrangian Monte Carlo tracer particles during the development of the Kelvin-Helmholtz instability (KHI) in 2D. The initial condition for this test problem is from \citet{2016MNRAS.455.4274L} and uses $P_0 = 10$, shear velocity of $1$, width of shear profile $\sigma = 0.2$, $a = 0.01$. The fluid solution is computed using the LLF Riemann solver with PLM spatial reconstruction and the RK2 temporal integrator within a 2D domain with resolution $512 \times 1024$ for $-0.5 < x < 0.5$ and $-1 < z < 1$. The passive contaminant is initialized according to (8e) of \citet{2013MNRAS.435.1426G}; and its state is shown at times $t=2$ and $t=3$ as computed by the fluid equations (top row) and as recovered from the Lagrangian tracer particles (bottom row). 

While the Lagrangian Monte Carlo method more accurately follows the mass flow and better recovers the statistical properties of the underlying fluid, it is more diffusive than the fluid evolution itself. The extra diffusivity is a fundamental consequence of the probabilistic update algorithm, results in an effective random walk with respect to the true history of the particles, and cannot be mitigated by increasing the number of particles. In practice, this means that the tracer particle histories may be interpreted as being noisy (see \citealt{2013MNRAS.435.1426G}).

\subsection{Test particles}

\begin{figure*}[htb]
    \centering
    \includegraphics[width=\linewidth]{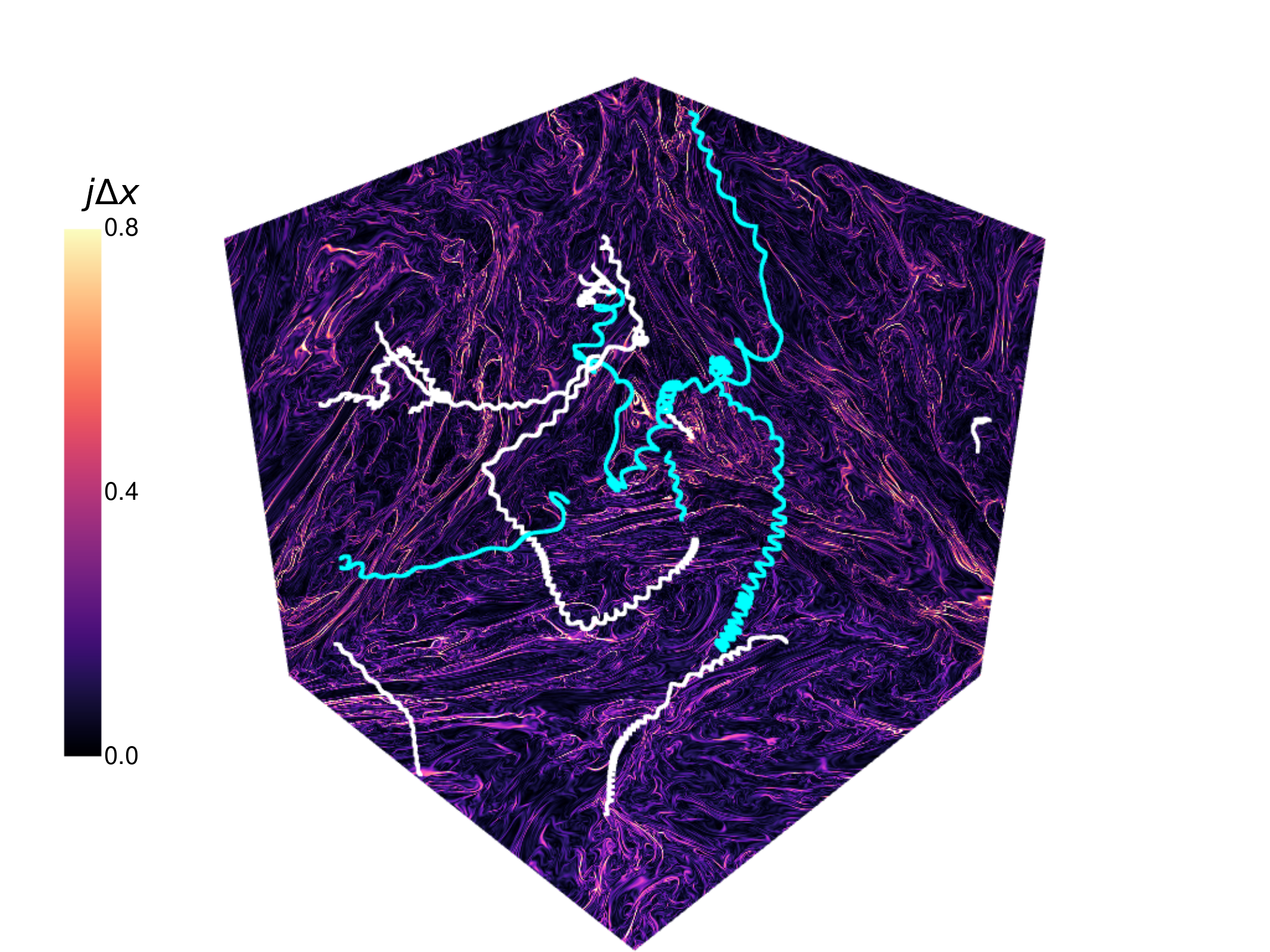}
    \caption{Current density (background colors) and charged particle trajectories for two different charge-to-mass ratios (cyan and white lines) in a 3D simulation of driven MHD turbulence at a resolution of $1280^3$.}
    \label{fig:particle-CR}
\end{figure*}

For many problems it is of great interest to follow the motion of charged test particles in the magnetic field produced by an MHD flow, for example to understand cosmic ray scattering by turbulence in the ISM, or the acceleration of particles in accretion disk turbulence.  In \ak{} such motion can be computed either from static snapshots of the flow in a post-processing mode, or dynamically by evolving both the MHD and particle motion simultaneously to capture the effect of intermittent structures. The former is well motivated in the study of cosmic ray scattering by inter-stellar turbulence, as typical inter-stellar gas velocities are several orders of magnitude smaller than the speed of light. In the latter case the ratio of the velocity of the particles compared to the fluid must be limited.

To integrate the motion of the particles, various pusher functions have been implemented in \ak{}, including a method based on the Boris algorithm to conserve the magnetic moment of particles \citep{Boris1970}, and a method for orbital integration in general relativistic curved spacetimes \citep{bacchini2019ApJS..240...40B}. The MHD fields are evaluated at the particle positions from the grid using either a linear interpolation method or the Triangular Shaped Cloud interpolation method \citep{BirdsallLangdon1991}.

As an example, Figure \ref{fig:particle-CR} shows the motion of select particles in isothermal, driven MHD turbulence on a $1280^3$ grid with a Reynolds number of $5 \times 10^4$ and a magnetic Prandtl number of 1. The turbulence simulation uses an HLLD Riemann solver and the same driving and spatial reconstruction as the $4096^3$ turbulence simulation described in \S3.2 and Figure \ref{fig:MHD_app}. After $\approx 7$ outer-scale eddy turnover times with steady-state $\delta B / \bar{B} \approx 3$, the turbulent field was restarted in static post-processing mode and $\approx 3.6 \times 10^7$ charged-particle trajectories were integrated for over 6 additional box-crossing times. The white and cyan lines in Figure \ref{fig:particle-CR} show example particle trajectories for two different charge-to-mass ratios (white lines corresponding to twice the charge-to-mass ratio of the cyan lines). The background shows the magnitude of the current density normalized by the grid spacing in three 2D slices through the 3D domain, highlighting the intermittency of the fields through which the particles propagate. 

\section{Performance and Scaling}

A primary motivation for developing \ak{} has been to use \kokkos{} to enable performance portability, and it is important to test if this goal has been achieved.  In this section we present the results of performance and weak scaling tests on various architectures.

\subsection{Single-Device Performance}

\begin{deluxetable*}{lcccccccc}
\tabletypesize{\footnotesize}
\tablecolumns{9} \tablewidth{30pt}
\tablecaption{Performance of \ak{} on various devices\label{table:perf-devices}}
\tablehead{
\colhead{\begin{tabular}{l} {} \\ {} \end{tabular}} &
\colhead{\begin{tabular}{c} Apple\\ M1 pro\\ (8 cores)  \end{tabular}} &
\colhead{\begin{tabular}{c} Intel Xeon\\ Gold 6326\\ (32 cores)\end{tabular}} &
\colhead{\begin{tabular}{c} NVIDIA\\V100   \end{tabular}} &
\colhead{\begin{tabular}{c} NVIDIA\\A100   \end{tabular}} &
\colhead{\begin{tabular}{c} AMD\\ MI250  \end{tabular}} &
\colhead{\begin{tabular}{c} NVIDIA\\ H100  \end{tabular}} &
\colhead{\begin{tabular}{c} NVIDIA\\ Grace Hopper  \end{tabular}}
}
\startdata
hydro    &34  &63  &298 &614  &405 & 677 & 1077 \\ 
MHD      &11  &33  &158 &298  &190 & 290 & 565
\enddata
\label{table:different-devices}
\end{deluxetable*}

Table \ref{table:perf-devices} shows the performance on a variety of devices (measured in cell-updates-per-device-second) of the Newtonian hydrodynamics and MHD solvers running a linear wave convergence test on a uniform mesh.  On CPUs, the mesh consists of a single $128^3$ {\tt MeshBlock}, and the test is performed using all cores available on a given CPU. Since the performance of \ak{} on CPUs is typically memory bandwidth limited, this represents a more realistic measurement appropriate to actual applications. On GPUs, we use a $512\times256\times256$ grid tiled with $128^3$ {\tt MeshBlocks}. The table demonstrates \ak{} is achieving excellent performance across the full range of devices tested.  On CPUs, the performance is comparable to that reported for \app{} (S20).  The latter included agressive vectorization of loops using compiler directives, while in \ak{} we rely on \kokkos{} \texttt{cmake} options to enable these options.  It is likely that further optimizations of \ak{} targeting CPUs is possible.  On GPUs, the performance shown in Table \ref{table:perf-devices} is comparable to, or better than, other GPU hydrodynamic and MHD codes reported in the literature, for example \texttt{CholloMHD} \citep{caddy2024ApJ...970...44C} or \texttt{AthenaPK} \citep{Holmen2024}.  On the latest generation of devices, the NVIDIA Grace Hopper processor, \ak{} achieves more than $10^9$ cell updates per second per device for 3D hydrodynamics.

As discussed in S20, we find the size of the {\tt MeshBlock}s has a significant effect on the measured performance.  For example, on an NVIDIA A100, reducing the {\tt MeshBlock} size to $64^3$, $32^3$ or $16^3$ while keeping the overall grid the same reduces performance from 614 Mzc/s to 413, 242 and 126 Mzc/s respectively.  Precisely the same behavior was observed on CPUs using \app{}, albeit for smaller {\tt MeshBlock} sizes.  It is clear that for best performance it is important to tune the size of the problem to the capability of the target device.  Running small problems on GPUs is not an effective use of resources.

\begin{deluxetable*}{lccccccc}
\tabletypesize{\footnotesize}
\tablecolumns{8} \tablewidth{30pt}
\tablecaption{Performance of various solvers in \ak{} on selected devices\label{table:perf-solvers}}
\setlength{\tabcolsep}{+0.05cm}
\tablehead{
\colhead{\begin{tabular}{l} {} \\ \end{tabular}} &
\colhead{Hydro} &
\colhead{MHD} &
\colhead{SR-hydro} &
\colhead{SR-MHD} &
\colhead{GR-hydro} &
\colhead{GR-MHD} &
\colhead{rad-GR-hydro}
}
\startdata
Intel Xeon Gold 6326  &63   &33  &16  &20  &13  &12  &6.2 \\ 
NVIDIA A100           &614  &298 &378 &178 &281 &120 &109 \\
NVIDIA GH             &1077 &565 &653 &297 &445 &227 &211
\enddata
\label{table:different-physics}
\end{deluxetable*}

To investigate the relative performance of the different physics solvers,
Table \ref{table:perf-solvers} shows the performance of each on selected devices.  For non-relativistic problems the performance is measured using the same linear wave convergence test as used for Table \ref{table:perf-devices}, while for SR and GR a 3D shocktube problem is used.   For non-relativistic, SR, and GR fluids the MHD solver runs roughly $2\times$ slower than the equivalent hydrodynamic solver on most devices, due to the extra complexity of the Riemann solvers and the cost of integrating the induction equation with CT.  The exception is for SR hydrodynamics on CPUs, which runs slower than SRMHD.  In SR and GR, the nonlinear root finding algorithm required to compute the primitive variables can add significant cost. Likely this algorithm requires further tuning on CPUs.  Again, \ak{} shows performance similar to other codes published in the literature, for example reaching 120 Mzcps for GRMHD on an NVIDIA A100, which is comparable to the performance reported for the highly optimized H-AMR code \citep{liska2022ApJS..263...26L}.

\begin{figure*}[htb]
    \centering
    \includegraphics[width=3.in]{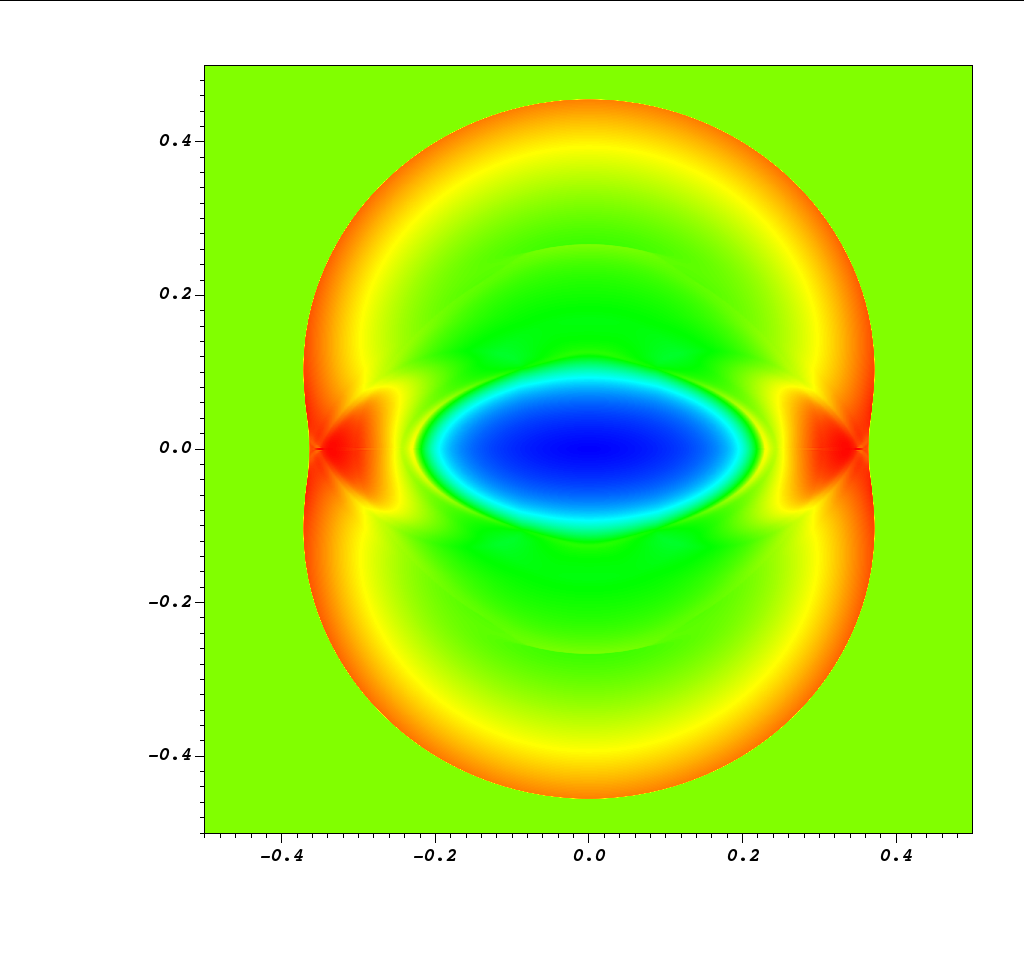}
    \includegraphics[width=3.in]{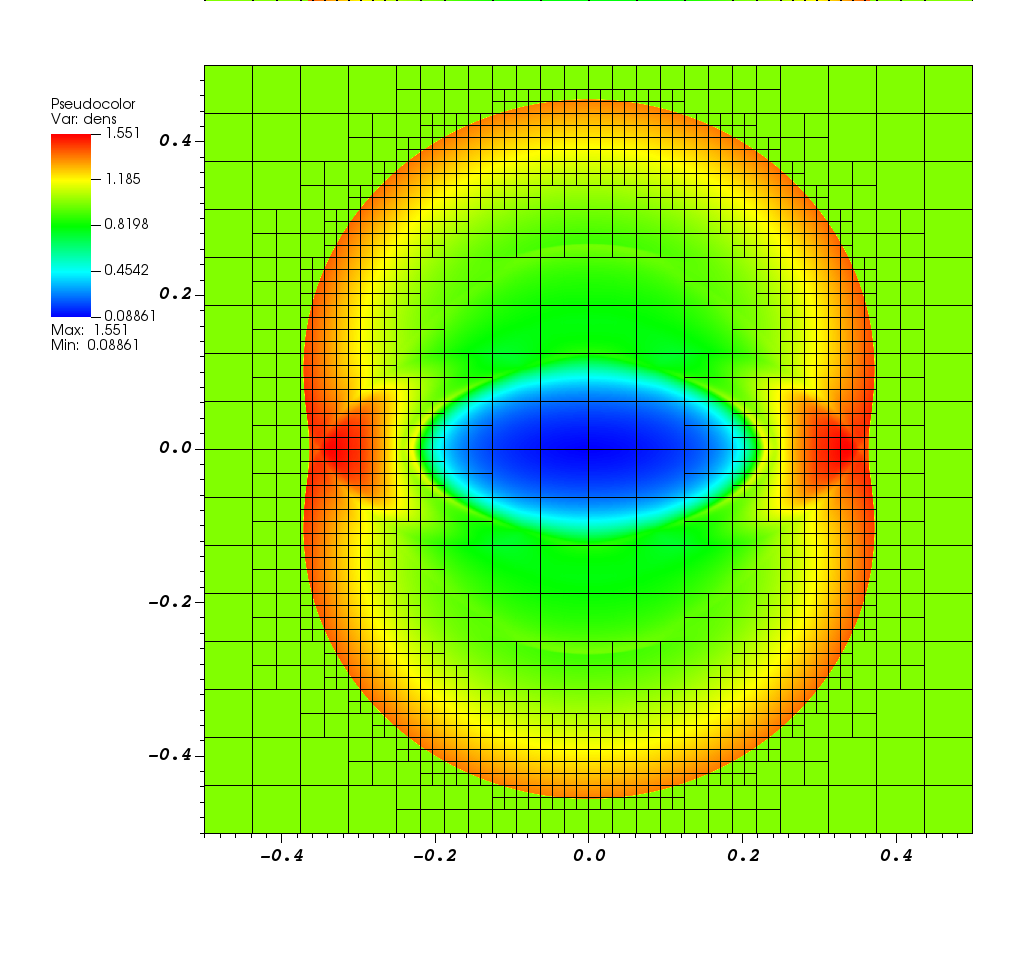}
    \caption{Density at time 0.4 in a 2D MHD blast test on a uniform (left) and AMR (right) grid run on a single NVIDIA A100 GPU.  Both calculations have the same effective resolution and produce identical results, however the AMR calculation is $5.4\times$ faster.}
    \label{fig:blast_amr}
\end{figure*}

An important capability of \ak{} is the AMR framework, and to demonstrate that the block-based AMR strategy adopted in the code can be run effectively 
on heterogeneous systems such as GPUs, we have run a 2D MHD blast wave test on a single NVIDIA A100 GPU.  The initial conditions consist of a uniform background with constant density and pressure to which a circular region with pressure $100\times$ the background is added. The test is run with both a uniform $4096^2$ grid with $64^2$ {\tt MeshBlocks}, as well as with AMR on a $1024^2$ root mesh with $64^2$ {\tt MeshBlocks} and a maximum of two levels of refinement (so as to achieve the same resolution as the uniform mesh).  Refinement occurs wherever the density exceeds the background by a factor of 1.1. Figure \ref{fig:blast_amr} shows that the resulting density profiles are essentially identical, confirming the fidelity of the AMR algorithm. Most importantly, using block-based AMR on GPUs only incurs a roughly 20\% overhead: the uniform mesh calculation runs at 181 Mzcps, while the AMR calculation runs at 141 Mzcps. In fact this overhead is likely amplified by the use of a 2D mesh, which significantly reduces the work associated with each {\tt MeshBlock} (a 3D version of this test is too large to fit on a single device). Nevertheless, the AMR calculation is significantly more efficient than using a uniform mesh, requiring only 18\% of the wall clock time.  This reduction results both because of the smaller number of {\tt MeshBlocks} needed with AMR and a larger timestep due to the coarser grid where the wavespeeds are maximal.

Overall, we find the AMR framework implemented in \ak{} using \kokkos{} achieves comparable performance to similar algorithms implemented in vendor specific programming models.  We conclude \kokkos{} has indeed enabled performance portability.

\subsection{Weak Scaling}

\begin{figure*}[htb]
    \centering
    \includegraphics[width=3.in]{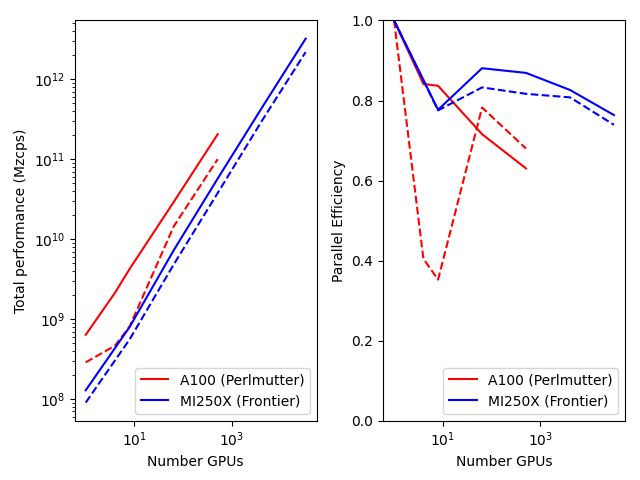}
    \caption{(Left) Total performance. (Right) Parallel efficiency.}
    \label{fig:weakscaling}
\end{figure*}

To test the parallel scaling of \ak{} on modern exascale HPC systems, we have performed weak scaling tests on both Perlmutter at NERSC (composed of NVIDIA A100s) and Frontier at OLCF (composed of AMD MI250X). Figure \ref{fig:weakscaling} shows the total performance (in Mzcps) and parallel efficiency (defined as performance per device compared to the performance on a single device) for both hydrodynamic and MHD linear wave convergence tests run on both machines. On Perlmutter there is a significant drop in performance when using a small number of nodes, however for larger jobs the parallel efficiency recovers to roughy 70\% on 4096 GPUs.  We note that both machines are in production mode with a heavy workload, and this likely affects the way that small jobs are distributed on the network.  On Frontier, \ak{} shows excellent weak scaling with a parallel efficiency of 80\% on up to 65536 GPUs (in this case, single GCDs).  We report similar results for the Z4c and dynamical GRMHD solvers presented in \cite{Zhu.9.24} and \cite{Fields.9.24} respectively.

\begin{figure*}[htb]
    \centering
    \includegraphics[width=3.in]{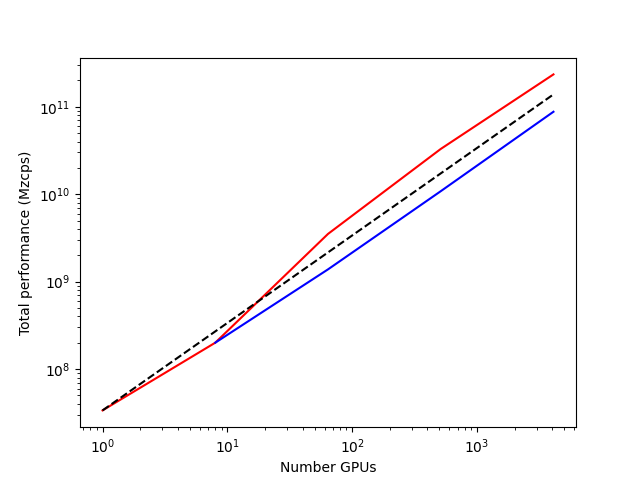}
    \caption{Total performance with AMR}
    \label{fig:amr_perf}
\end{figure*}

To test the parallel efficiency of the block-based AMR framework, we present a weak scaling test of the propagation of a linear sound wave in non-relativistic hydrodynamics.  On a single device, a $128^3$ root grid is decomposed into $16^3$ {\tt MeshBlock}s with one level of mesh refinement wherever the mass density is larger than the background by 50\% of the wave amplitude.  The wave propagates along the diagonal of the mesh, which introduces a complex pattern of refined {\tt MeshBlock}s that must be continuously refined and de-refined as the wave propagates.  On multiple devices, load balancing is required to keep the work per device roughly constant.  Figure \ref{fig:amr_perf} shows the resulting total performance for a test run on Frontier at OLCF on up to 4096 GPUs (GCDs). Two different tests were run.  The red line shows the performance as both the mesh and {\tt MeshBlock} size is increased (the latter from $16^3$ on one device to $128^3$ on 512 devices, after which it is also held constant.  The blue line shows the results as the {\tt MeshBlock} size is held fixed at $32^3$ above 8 GPUs.  In the former case, super-linear scaling is observed, as the efficiency increases with increasing {\tt MeshBlock} size.  However, even for fixed {\tt MeshBlock} size (blue line) nearly perfect weak scaling is observed, even though at the highest resolution over 5 million {\tt MeshBlock}s were communicated between nodes in order to achieve good load balancing.  It is clear the AMR algorithms in \ak{} are indeed well suited to modern HPC architectures.

\section{Discussion and Summary}

In this paper we have described \ak{}, a new implementation of the \app{} AMR framework and various astrophysical fluid dynamic, radiation transport, and particle kinetic solvers using \kokkos{}. In two companion papers, we describe a numerical relativity solver \citep{Zhu.9.24} and a GRMHD solver in dynamical spacetimes \citep{Fields.9.24} based on the \ak{} framework. We have demonstrated that \ak{} achieves excellent performance and weak scaling across a wide range of architectures.  On 65536 GPUs on the OLCF Frontier system, the non-relativistic hydrodynamic and MHD solvers both achieve a parallel efficiency of roughly 80\%.  Moreover we have shown that the block-based AMR strategy adopted by both \app{} and \ak{} is well suited for modern exascale computing systems, achieving nearly perfect weak scaling for a hydrodynamic linear wave test. The code is therefore capable of leveraging modern exascale computing systems for solving challenging problems in astrophysics.

\ak{} is a small code designed to solve specific applications at exascale.  Thus, it does not implement the very broad range of algorithms and physics options that are included in \app{} (e.g., curvilinear coordinates). Given that it is much easier to maintain and update a smaller code, it is likely that most of the features of \app{} will not be implemented in \ak{} anytime soon.  Of course, it is still possible to adopt \app{} or other similar codes when a particular application requires such features.  Nevertheless there are several directions for future developments that are already in progress or planned for the near future.  These include the addition of the fourth-order integration algorithm of \cite{felker2018JCoPh.375.1365F}, self-gravity on the AMR mesh using multigrid \citep{tomida2023ApJS..266....7T}, and non-relativistic radiation and neutrino transport \citep{jiang2012ApJS..199...14J}.

While this paper demonstrates that the fundamental numerical algorithms adopted in \ak{} are well suited to the more complex design of modern HPC systems, it is nevertheless becoming increasingly challenging to develop and maintain software as part of large research projects on such systems. A significant challenge for the future will be finding the resources and expertise to develop and support increasingly complex research software.  It is likely that open source frameworks such as \kokkos{} and \parth{} will be important in this effort.

\acknowledgments

We thank Kengo Tomida for discussion regarding the AMR framework, and the many other contributors to the \app{} code project. We thank the members of the \parth{} collaboration whose input on programming with \kokkos{} was essential.
We thank Matthew Cawood for providing performance numbers on NVIDIA's Grace Hopper processor. 

JS was supported by a subcontract from the Texas Advanced Computing Center from National Science Foundation grant 2139536, and by the Eric and Wendy Schmidt Funds for Strategic Innovation. 
ERM acknowledges partial support
by the National Science Foundation under grants No. PHY-
2309210, AST-2307394, and from NASA under grant 80NSSC24K1229.

Simulations were performed on OLCF's Frontier, ALCF's Polaris, NERSC's Perlmutter, and on Princeton University's Della and Stellar supercomputers.

Many figures in this manuscript were produced with \texttt{matplotlib} \citep{Hunter:2007}, the \texttt{yt} package \citep{2011ApJS..192....9T}, or the \texttt{VisIt} visualization software \citep{Childs_High_Performance_Visualization--Enabling_2012}.

This research used resources of the National Energy Research Scientific
Computing Center, a DOE Office of Science User Facility supported by the Office of Science of the U.S.~Department of Energy under Contract No.~DE-AC02-05CH11231.  

This research used resources of the Oak Ridge Leadership Computing Facility at the Oak Ridge National Laboratory, which is supported by the Office of Science of the U.S. Department of Energy under Contract No. DE-AC05-00OR22725.

This research used resources of the Argonne Leadership Computing Facility, a U.S. Department of Energy (DOE) Office of Science user facility at Argonne National Laboratory and is based on research supported by the U.S. DOE Office of Science-Advanced Scientific Computing Research Program, under Contract No. DE-AC02-06CH11357.

The simulations presented in this article were partly performed on computational resources managed and supported by Princeton Research Computing, a consortium of groups including the Princeton Institute for Computational Science and Engineering (PICSciE) and the Office of Information Technology's High Performance Computing Center and Visualization Laboratory at Princeton University. 

Research presented in this article was supported by the Laboratory Directed Research and Development program of Los Alamos National Laboratory under project number 20220087DR.

This work has been assigned a document release number LA-UR-24-30139.

\newpage

\bibliography{main}{}
\bibliographystyle{aasjournal}

\end{document}